\documentclass[a4paper,fleqn,usenatbib]{mnras}

\usepackage{newtxtext,newtxmath}

\usepackage[T1]{fontenc}
\usepackage{ae,aecompl}

\usepackage{graphicx}	
\usepackage{amsmath}	
\usepackage{amssymb}	
\usepackage{subfig}
\usepackage{enumitem}
\usepackage{color}
\usepackage[dvipsnames]{xcolor}

\newcommand{\premise}{\texttt{premise}}
\usepackage[T1]{fontenc} 
\usepackage{aecompl}

\title[Sparse parameter estimation]{Sparse estimation of model-based diffuse thermal dust emission}

\author[M. O. Irfan et al.]{
Melis O. Irfan,$^{1}$\thanks{E-mail: melis.irfan@cea.fr}
J\'er\^ome Bobin,$^{1}$
\\
$^{1}$Laboratoire CosmoStat, AIM, UMR CEA-CNRS-Paris 7 Irfu, SAp/SEDI, Service d'Astrophysique, \\ CEA Saclay, F-91191 GIF-SUR-YVETTE CEDEX, France.\\
}

\date{Accepted XXX. Received YYY; in original form ZZZ}

\pubyear{2017}

\begin{document}
\label{firstpage}
\pagerange{\pageref{firstpage}--\pageref{lastpage}}
\maketitle

\begin{abstract}
Component separation for the {\it{Planck}} HFI data is primarily concerned with the estimation of thermal dust emission, which requires the separation of thermal dust from the cosmic infrared background (CIB). For that purpose, current estimation methods rely on filtering techniques to decouple thermal dust emission from CIB anisotropies, which tend to yield a smooth, low-resolution, estimation of the dust emission. In this paper we present a new parameter estimation method, {\texttt{premise}}: Parameter Recovery Exploiting Model Informed Sparse Estimates. This method exploits the sparse nature of thermal dust emission to calculate all-sky maps of thermal dust temperature, spectral index and optical depth at 353\,GHz. {\texttt{premise}} is evaluated and validated on full-sky simulated data. We find the percentage difference between the {\texttt{premise}} results and the true values to be 2.8, 5.7 and 7.2 per cent at the 1$\sigma$ level across the full sky for thermal dust temperature, spectral index and optical depth at 353\,GHz, respectively. Comparison between {\texttt{premise}} and a GNILC-like method over selected regions of our sky simulation reveals that both methods perform comparably within high signal-to-noise regions. However outside of the Galactic plane  {\texttt{premise}} is seen to outperform the GNILC-like method with increasing success as the signal-to-noise ratio worsens.
\end{abstract}




\section{Introduction}

Within our Galaxy dust grains, such as carbonaceous and amorphous silicates \citep{draine}, are heated by inter-stellar, UV radiation. The resulting emission, known as thermal dust emission, is the dominant diffuse Galactic emission at frequencies{\mbox{ $>$ 100\,GHz}} \citep{bennett}. Measurements of thermal dust emission reveal the chemistry of the interstellar medium (\citet{comp11}; \cite{jones}), trace interstellar radiation and reveal dust mass and dust column density. Additionally, thermal dust emission is a prominent astrophysical foreground present in measurements of the cosmic microwave background (CMB). In order to infer cosmological information from measurements of the CMB over 100\,GHz, accurate subtraction of thermal dust emission is crucial \citep{planckCompSep}.

Over large angular-scales thermal dust emission can be modelled as blackbody emission, as can the cosmic infrared background (CIB); diffuse infrared emission from dusty galaxies across all redshifts. The CIB is a strong tracer of the star formation history of the Universe \citep{lagache} and its anisotropies shed light on galaxy clustering and dark matter halo distributions \citep{bethermin}. The similarity in spectral shape between thermal dust emission and the CIB adds complexity to their separation. The correlation between thermal dust and $\rm{H_{I}}$ emission can be used to differentiate between thermal dust and the CIB at large angular scales \citep{pr2} but unfortunately this correlation cannot be exploited at scales smaller than several degrees. 

Since the second {\it{Planck}} data release specific attempts to separate thermal dust emission from the CIB over {\it{Planck}} HFI frequencies have been implemented with increasing success. \citet{pr2} fit a modified blackbody (MBB) to the data and present all-sky maps of the three fit parameters: dust temperature, spectral index and optical depth at 353\,GHz. The MBB parameters do not represent the mass-weighted averages of the physical dust properties along the line-of-sight, e.g. the MBB dust temperature is not the mass-weighted temperature of all the dust particles contributing to the measured emission. Rather, the MBB parameters are simply the empirical parameters which best fit the overall dust emission. This approach is well-suited for component separation, however for an improved understanding of the thermal dust emission mechanism itself more complex models are required (e.g. \citet{draine}; \citet{comp11}; \citet{jones}).

\citet{pr2} introduce a two-stage approach to the customary pixel-by-pixel fit. First they smooth the $N_{\rm{side}}$ 2048 HFI maps to 30 arcmin and fit a MBB with three free parameters to obtain the spectral index per pixel. Next they fit the 5 arcmin (essentially full resolution) HFI maps, fixing the spectral index values to those obtained at 30 arcmin and allowing two free parameters for the fit. The result being all-sky maps of dust temperature and 353\,GHz optical depth at 5 arcmin and spectral index at 30 arcmin. Smoothing the data averages out the CIB anisotropies to a greater extent than it averages out the spatial correlations of thermal dust emission; the CIB has a flatter angular power spectrum than thermal dust. However the effects of smoothing on thermal dust emission are not negligible; \citet{pr2} note the trade-off between sensitivity to small-scale variations of thermal dust emission and smoothing for component separation purposes. The all-sky maps of thermal dust emission formed from the three fitted parameters display variations of 30 per cent, at all scales, from the Finkbeiner, Davis and Schlegel thermal dust model \citep{fds}. 

\citet{gnilc} present an improvement on the \citet{pr2} method by only smoothing the data within the regions where the CIB anisotropies start to dominate the total signal at small angular scales. Their analysis is performed within the needlet (spherical wavelet) domain, where emissions are modelled as transient waveforms within pixel and harmonic space \citep{gnilc11}. \citet{gnilc} account for the `nuisance'  (non-thermal dust) contributions using a covariance matrix formed from noise, CIB and CMB estimates. This nuisance covariance matrix is used to identify regions of the sky where thermal dust emission dominates over CIB, CMB and noise. Within the nuisance dominated regions the data are smoothed and reassessed, if the nuisance term continues to dominate the width of the Gaussian kernel used for smoothing is increased by a factor of two and the process repeated. Therefore the \citet{gnilc} all-sky estimates of thermal dust emission have a range of effective beam sizes from FWHM = 5.0 arcmin, within the high signal-to-noise regions (over 65 per cent of the sky) to FWHM = 21.8 arcmin at high Galactic latitudes. The \citet{gnilc} thermal dust maps are shown to be an improvement on the \citet{pr2} maps when the residual CIB maps from both methods and the reduced $\chi^2$ from the MBB fits are compared. 

\subsection{Limitations and objectives}

Both the \citet{pr2} and \citet{gnilc} method have the disadvantage of being computationally intensive, as they perform the MBB fit on each pixel for the full resolution {\it{Planck}} data (over fifty million pixels). While the \citet{pr2} method has the ability to provide thermal dust temperature and 353\,GHz optical depth maps at the full 5 arcmin resolution, any variations in the dust spectral index at angular scales smaller than 30 arcmin are smoothed over. Contrarily though, it could be argued that the \citet{pr2} maps are not smoothed enough as strong evidence of CIB contamination still remain. Therefore it is clear that smoothing all pixels within the full sky HFI maps to the same extent is not optimum. The strength of the \citet{gnilc} method is that it locates the areas worst affected by CIB contamination and applies the largest degree of smoothing within these areas. As successful as this method is it still presents an incomplete picture of the spatial variance of the dust parameters across the sky, due to smoothing. We aim to provide a method of parameter estimation for thermal dust emission, contaminated by CIB and noise contributions, which can be applied to all-sky data at full {\it{Planck}} resolution, is robust to a wide range of signal-to-noise ratios across the sky and is faster than a pixel-by-pixel MBB fit. 

We present a new dust parameter estimation method, which we refer to as {\texttt{premise}}: Parameter Recovery Exploiting Model Informed Sparse Estimates. This novel approach builds upon a two-stage procedure:
\begin{itemize}
\item a filtering technique that first performs CIB and noise removal. In the spirit of GNILC, it makes use of the nuisance covariance matrix but further exploits the natural sparsity of the dust emission in the wavelet domain.\\

\item an innovative parameter estimation algorithm that builds upon recent advances in applied mathematics to find the optimum parameter values per pixel. This is achieved using the least squares estimator on the residual between the model maps and the empirical data (thermal dust plus nuisance terms) within the wavelet domain and adding a penalisation factor to favour sparsity.
\end{itemize} 
As {\texttt{premise}} never runs a pixel-by-pixel MBB fit it is faster than the traditional methods and the use of sparsity in the second stage reduces the need for smoothing whilst still providing an improvement in parameter estimation within noise dominated regions. 

Starting with simulation data representing the combination of thermal dust emission, point sources, instrumental noise, CMB and CIB we apply {\texttt{premise}} to produce estimates of the thermal dust MBB parameters. The simulated thermal dust emission was created using a single MBB with three parameters: temperature, spectral index and optical depth at 353\,GHz. Fitting the same model to the {\texttt{premise}} thermal dust emission maps provides a comparison between the fitted and the true parameters which can be used to evaluate {\texttt{premise}}. Additionally, we implement a GNILC-like methodology \citep{gnilc} to enable comparisons between GNILC and {\texttt{premise}}.

\subsection{Notations}

Throughout this work we make use of numerous notations, we summarise them here for clarity. $x_{\nu_{i}}[k]$ indicates the total flux for pixel $k$ at frequency $\nu_{i}$ and is the combination of CMB, CIB and instrumental noise, point sources and thermal dust emission:
\begin{equation}
x_{\nu_{i}}[k] = x_{\nu_{i}}[k] ^{\rm{CMB}} + x_{\nu_{i}}[k] ^{\rm{CIB}} + x_{\nu_{i}}[k] ^{\rm{noise}} + x_{\nu_{i}}[k] ^{\rm{ps}} + x_{\nu_{i}}[k] ^{\rm{dust}}.
\end{equation}  
When considering all of the observational frequencies simultaneously the vector form of the above equation can be used:
\begin{equation}
X[k] = X[k] ^{\rm{CMB}} + X[k] ^{\rm{CIB}} + X[k] ^{\rm{noise}}  + X[k] ^{\rm{ps}} + X[k] ^{\rm{dust}}
\end{equation}  
and for all frequencies and all pixels we have the matrix form:
\begin{eqnarray}
{\bf{X}} = {\bf{X}} ^{\rm{CMB}} + {\bf{X}} ^{\rm{CIB}} + {\bf{X}} ^{\rm{noise}} + {\bf{X}} ^{\rm{ps}} + {\bf{X}}^{\rm{dust}}.
\end{eqnarray} 
The forward wavelet transformation of single frequency map is denoted as  $x_{\nu_{i}} {\bf{\Phi}}$. At each scale the transformations produces $2^{j-1}$ coefficients which consist of the coarse scale ($c$) and wavelet coefficients ($w$):
\begin{eqnarray}
x_{\nu_{i}} \Phi  = \alpha_{\nu_{i}} = [c_{\nu_{i}}, w_{\nu_{i}}^{(j=1)} ... w_{\nu_{i}}^{(j=J)}],
\end{eqnarray}
with $0 < j < J$ being the number of the wavelet scale. The L1-norm of a variable in the wavelet domain is equivalent to the sum of the wavelet coefficients:
\begin{eqnarray}
 \| x_{\nu_{i}} {\bf \Phi}\|_{\ell_1} = \sum_j | w_{\nu_{i}}^{  j}|.
\end{eqnarray}
Formally, the wavelet decomposition of the data matrix $\bf X$ yields the following decomposition at each scale $j$:
\begin{eqnarray}
{\bf W}_j = {\bf W}_j^{\rm{CMB}} + {\bf W}_j^{\rm{CIB}} + {\bf W}_j ^{\rm{noise}}  +  {\bf W}_j ^{\rm{ps}} +  {\bf W}_j^{\rm{dust}}
\end{eqnarray}
where the rows of the matrix ${\bf W}_j$ contain the $j$-th wavelet scale of each of the channels. 

$\mathcal{S}_{\tau}$ stands for the soft-thresholding operator with threshold $\tau$, which is described as follows for some value $u$:
\begin{equation}
\mathcal{S}_{\tau}(u) = \left \{
\begin{array}{ll}
u - \tau \mbox{sign}(u) & \mbox{ if } |u| > \tau \\
0 & \mbox{ otherwise}
\end{array}
\right.
\end{equation}

When referring to an iterative process we denote a variable at a particular time `t' as $x^{(t)}$ and specifically for gradient descent iterations we define the minimum of said variable as $x^{\frac{1}{2}}$.  \\
 
The paper is organised as follows: section \ref{sec:data} introduces the simulation data alongside the ancillary data used, section \ref{sec:method} details the steps within {\texttt{premise}} and section \ref{sec:results} presents our results in comparison with results obtained via the GNILC-like methodology.  

\section{Data and preprocessing}
\label{sec:data}
\subsection{Simulated Thermal Dust}
The data used in this paper are simulation data, formed from equation~(\ref{eq:bbinten}).
The GNILC \footnotemark  all-sky temperature and spectral index maps provide the required temperature ($T$)
 and spectral index ($\beta$), the {\it{Planck}} FFP8 \citep{ffp} thermal dust map at 353\,GHz provides the normalisation factor and $B$ denotes the blackbody function.  
\footnotetext{http://pla.esac.esa.int/pla/}
\begin{eqnarray}
x_{\nu_{i}} ^{\rm{dust}} = x_{353\, \rm{GHz}}^{\rm{ffp8}} \times \frac{B(T,  \nu)}{B(T, 353 \, \rm{GHz})} \times
    \left(\frac{\nu}{353 \, \rm{GHz}}\right)^{\beta} \times \rm{cc}^{-1}.
\label{eq:bbinten}
\end{eqnarray}
The single spectral index model was preferred over the two-index model \citep{meis} as the two-index model has been shown to improve the fit to the data when frequencies below 353\,GHz are used. Equation~(\ref{eq:bbinten}) was evaluated at frequencies 353, 545, 857 and 3000\,GHz to emulate {\it{Planck}} HFI data used in combination with the 100\,$\mu$m IRIS map \citep{iris}. These fluxes were multiplied by their inverse colour corrections ($\rm{cc}^{-1}$) to further align them with real data. The colour corrections were calculated across the {\it{Planck}} HFI bandpasses for frequencies 353, 545 and 857\,GHz. The 3000\,GHz colour correction factors were taken from Table 3 of \citet{iris}. 

Simulation data were not created for the lowest two HFI frequencies (100 and 217\,GHz) following the method of \citet{gnilc}, which fits a MBB model to the GNILC 353, 545, 857 and 3000\,GHz thermal dust estimates. The 353\,GHz FFP8 data were neither smoothed nor downgraded, so all our simulation data are at $\rm{N_{side}}$ 2048 with a FWHM of $\sim$ 5 arcmin. 

\subsection{Simulated CIB, CMB, noise and point sources}
\label{sec:simCIB}

The simulated total flux maps used differ for the GNILC-like methodology and {\texttt{premise}}. While GNILC includes the CMB within its covariance matrix of nuisance terms, {\texttt{premise}} is intended for use on CMB subtracted maps. In \cite{lgmca}, we introduced a sparsity-based component separation method coined L-GMCA to estimate a precise CMB map from the {\it{Planck}} data. More specifically, we showed that the estimated CMB map has very low foreground contamination. Consequently, the L-GMCA CMB map will be removed prior to applying the \premise \, algorithm. 

The total flux maps used for the GNILC-like methodology were constructed as: 
\begin{eqnarray}
{\bf{X}}^{\rm{GNILC-like}} =  {\bf{X}}^{\rm{dust}} +  {\bf{X}}^{\rm{CIB}} +  {\bf{X}}^{\rm{CMB}}  +   {\bf{X}}^{\rm{noise}}  + {\bf{X}}^{\rm{ps}},
\label{eq:gnilcTotS}
\end{eqnarray}
where the FFP8 simulations provide the instrumental noise, point sources and CMB contributions. A 3000\,GHz CIB map of $\rm{N_{side}}$ 2048 and FWHM 5 arcmin was created using the methodology detailed in Appendix C of \citet{pr2}.  Gaussian noise with a median level of 0.06 MJy $\rm{sr}^{-1}$ \citep{pr2} was used for the 3000\,GHz noise map. No CMB nor point source contributions were included at 3000\,GHz to replicate the point source subtracted version of the IRIS data made available by the Planck collaboration \citep{pr2}.
The total flux maps used for {\texttt{premise}} were constructed as: 
\begin{eqnarray}
{\bf{X}}^{\rm{{\texttt{premise}}}} = {\bf{X}}^{\rm{dust}} + {\bf{X}}^{\rm{\widetilde{CIB}}} +  {\bf{X}}^{\tilde{n}}(\nu)  + {\bf{X}}^{\rm{ps}},
\label{eq:waveTotS}
\end{eqnarray}
where ${\bf{X}}^{\rm{\widetilde{CIB}}}$  and ${\bf{X}}^{\tilde{n}}(\nu)$ are simulated CIB and instrumental noise maps featuring fractional CMB emission, representative of the CIB and noise properties of an intensity map after the L-GMCA CMB subtraction.

\subsection{LAB data and the large-scale CIB contribution}

\begin{figure}
\centering
\includegraphics[width=0.8\linewidth]{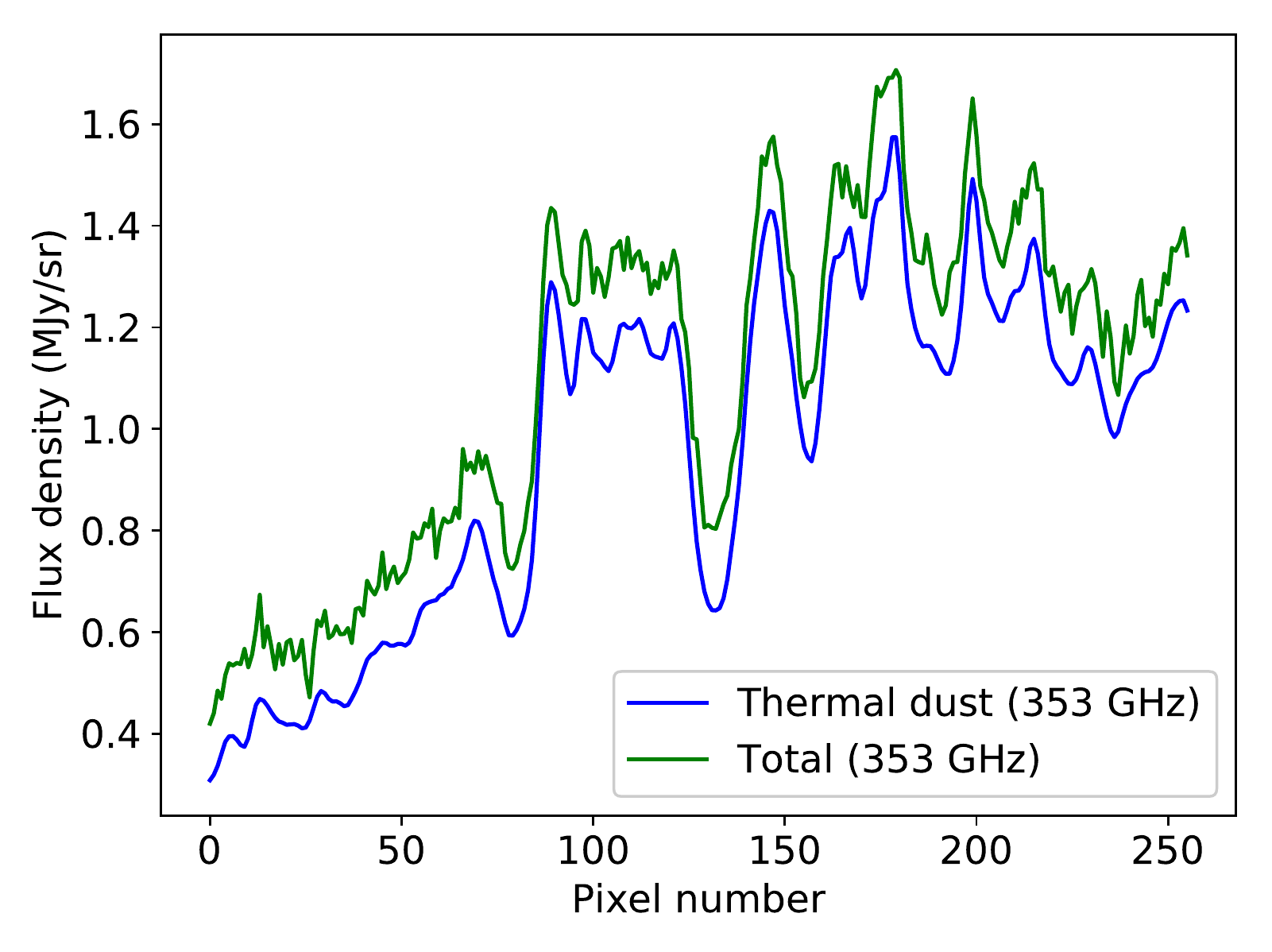}
\caption{One dimensional slice of a patch of data at 353\,GHz. The green line represents thermal dust plus CIB plus instrumental noise, the blue line is just thermal dust.}
\label{fig:ciboffset}
\end{figure}

The CIB is seen to contribute to the overall measured intensity in two ways: small-scale variations and a large-scale intensity which manifest as a constant, additive offset to the pure thermal dust intensity. Fig.~\ref{fig:ciboffset} shows a 1D slice of a patch of data at 353\,GHz. The green line shows the total emission while the blue line shows the pure thermal dust emission. The large-scale CIB contribution can clearly be seen as a positive offset while the small-scale CIB contributions, alongside instrumental noise, are seen as Gaussian-like variations around the thermal dust mean level. 

Before the simulated maps of total emission can be processed the large-scale CIB offsets must first be subtracted. The removal of the large-scale CIB offset is achieved through the use of LAB $\rm{H_{I}}$ Survey data \citep{lab}. These data are available in the form of all-sky {\texttt{HEALPix}} maps \citep{healpix} at $\rm{N_{side}}$ 512 and a FWHM of \mbox{$\sim$ 36 arcmin}. \footnotemark \footnotetext{https://lambda.gsfc.nasa.gov} The constant CIB offsets for each frequency were calculated using the method described in \citet{pr2}, see Appendix \ref{sec:apA} for details. The resulting offsets were found to be 0.126, 0.331, 0.641 and 0.657 MJy $\rm{sr^{-1}}$ for 353, 545, 857 and 3000\,GHz respectively.  These values were subtracted from the total flux maps. The mean values for the simulated CIB maps are 0.140, 0.366, 0.718 and 0.718 MJy\,sr$^{-1}$, placing an 8.5 -- 11 per cent error on the linear regression method.

\subsection{GNILC-like and {\texttt{premise}} preprocessing}

\begin{figure}
\centering
\includegraphics[width=0.99\linewidth]{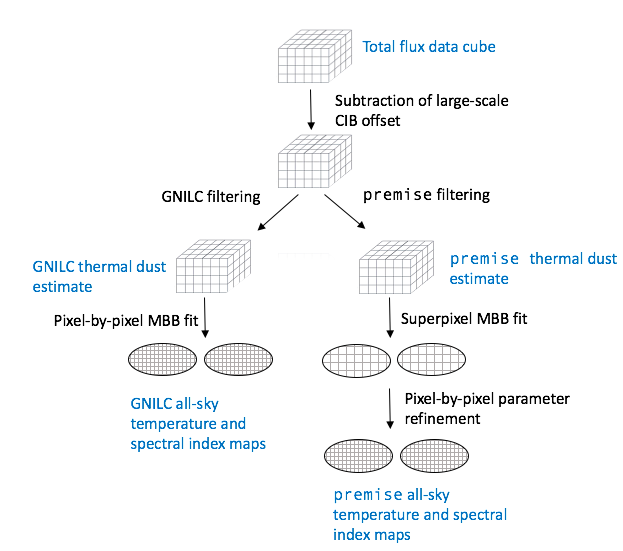}
\caption{The principal steps of the both the GNILC and {\texttt{premise}} algorithms.}
\label{fig:methFlo}
\end{figure}

We implement two algorithms: our own, {\texttt{premise}} and a GNILC-like methodology. The principal steps for both algorithms are summarised in Fig.~\ref{fig:methFlo}: removing large-scale CIB offsets, filtering out noise and small-scale CIB anisotropies and fitting a MBB. GNILC makes use of a covariance matrix of the combined nuisance terms to locate dimensions within signal subspace where the desired signal, thermal dust emission in this case, is dominant. The nuisance terms are the small-scale CIB contributions, the CMB and instrumental noise. As this paper is concerned with simulation data the exact CIB, CMB and instrumental noise maps used to make the total flux maps are given as nuisance term estimates, therefore the estimation of the nuisance covariance matrix is unrealistically perfect. The GNILC-like methodology for filtering total flux maps to produce pure thermal dust estimates that we implement is not identical to that described in \citet{gnilc}, therefore we detail our GNILC-like filtering technique in Appendix \ref{sec:apB}, highlighting any differences. 

After the initial filtering the following MBB model is fit pixel-by-pixel to the GNILC estimate of pure thermal dust emission: 
\begin{eqnarray}
x_{\nu_{i}} = \tau_{353} \times B(T,  \nu) \times \left(\frac{\nu}{353 \, \rm{GHz}}\right)^{\beta} \times \rm{cc}^{-1},
\end{eqnarray}
where the optical depth at 353\,GHz ($\tau_{353}$), the dust temperature ($T$) and the thermal dust spectral index ($\beta$) are the free parameters in the fit. The inverse colour correction ($\rm{cc}^{-1}$) also depends on $T$ and $\beta$.

From Fig.~\ref{fig:methFlo} it can be seen that both GNILC and {\texttt{premise}} filter the total flux maps to produce thermal dust estimates. For GNILC, the principal aim is to construct all-sky thermal dust estimates which can then provide values for the MBB parameters through a pixel-by-pixel fit. The goal of {\texttt{premise}} is the inverse; the {\texttt{premise}} filtered dust estimates are only of use as inputs to the MBB fit, our aim is to calculate the thermal dust MBB parameters as accurately as possible. 

\subsection{Dealing with point sources} \label{scontrib}

Point sources were removed from the 353, 545 and 857\,GHz data using the appropriate FFP8 masks for each frequency. The {\texttt{premise}} algorithm is capable of processing masked data as the use of the discrete wavelet transform allows for the reconstruction of the full wavelet, providing the Shannon sampling condition is fulfilled. The GNILC methodology, however, requires each data pixel to provide astrophysical information. Therefore the masked GNILC-like data were inpainted using a morphological component analysis technique \citep{inpaint} applied to each frequency map independently.   

\section{The\, \premise \,method}
\label{sec:method}

Estimating the thermal dust parameters per pixel from the {\it{Planck}} data raises two important issues: i) it requires solving large-scale ({\it i.e.} over fifty million pixels for full-resolution {\it{Planck}} data) non-linear parameter estimation problems and ii) it is highly sensitive to nuisance terms such as noise, CIB and CMB, which explains why standard pixel-based fitting performs poorly in noise dominated regions of the sky. The method of filtering nuisance contributions prior to applying pixel-based fitting has two limitations: i) it only uses statistical information about the nuisance term but does not account for the statistics of the dust to be retrieved, and ii) pixel-based fitting will still be sensitive to dust estimation biases induced by the dust filtering procedure. For that purpose, the proposed\, \premise \,method is a new parameter estimation procedure that minimises a non-linear least-square cost function to which an extra penalisation is added to favour sparse parameter maps in the wavelet domain. This procedure allows for a fast and effective full-sky estimation and accounts for the naturally sparse distribution of the temperature and spectral index maps in the wavelet domain.

\citet{pr2} split their MBB fit into two stages to combat parameter degeneracies introduced by the CIB anisotropies, choosing first to calculate the thermal dust spectral indices and then the temperature and optical depth at 353\,GHz. We too experienced such degeneracies and so also opt for a two stage approach, choosing to first calculate $T$ and $\beta$ and then, through the introduction of a dust template, we attempt to break the parameter degeneracies and calculate $\tau_{353}$.

\subsection{Fitting $T$ and $\beta$ from the MBB model}
Fitting for the spectral index and the temperature in the MBB model requires dealing with the non-linear relationship between these parameters and the observed pixel. For a given pixel, this reads as:
\begin{eqnarray}
\min_{\beta,T}  \sum_{i} \left({x}_{\nu_i}[k] - { x}_{\nu_i}[k]^{\mbox{dust}}(\beta[k],T[k]) \right)^2
\label{eq:pixfit}
\end{eqnarray}
In\, \premise \,and unlike currently available pixel-based fitting methods, we propose estimating full-sky maps of the parameters while accounting for their strong correlation across pixels. This further requires solving a minimisation problem of the form:
\begin{equation}\label{eq:nl_sparseLS}
\begin{split}
\min_{\beta,T} & \lambda_b \| \beta {\bf \Phi}\|_{\ell_1} +  \lambda_t \| T {\bf \Phi}\|_{\ell_1} + \\ & \sum_{i,k} \left({x}_{\nu_i}[k] - \mathcal{M}_{\nu_i}[k] { x}_{\nu_i}[k]^{\mbox{dust}}(\beta[k],T[k]) \right)^2
\end{split}
\end{equation}
where the quadratic term measures the discrepancy between the observations and the MBB model. Point source masks ($\mathcal{M}_{\nu_i}[k]$) take the value of $1$ when no source is present and $0$ otherwise. The penalisation terms enforce the sparsity of the estimated $T$ and $\beta$ in the wavelet domain through the sum of the wavelet coefficients. In the next section, we will make use of the undecimated and isotropic wavelet transform \citep{starck2007undecimated}.

We will show in section~\ref{parref} how such a penalised non-linear least-square minimisation problem can be solved using an iterative thresholding algorithm. However, since the MBB model is non-linear, these algorithms can be sensitive to the initial point from which the algorithm starts. To alleviate this issue, we propose a two-stage fitting approach that is composed of: i) a fast initialisation procedure that builds upon a quadtree decomposition of the data in the wavelet domain, and ii) a refinement stage based on an iterative thresholding algorithm.

\subsubsection{Fast wavelet-based initialisation}

The goal of this initialisation step is to provide a first estimate of the dust emission that is robust to CIB and noise, with low computational cost. For that purpose, and similarly to GNILC, a model fitting is performed on CIB and noise filtered data. In contrast to pixel-by-pixel fitting, the proposed initialisation procedure makes use of the quadtree technique, which significantly reduces the computational cost of this procedure. This eventually yields a trade-off between good estimation accuracy and robustness to noise and CIB but to a lesser extent than smoothing.\\

\paragraph*{Disentangling between the dust emission and nuisance contamination}
 
The filtering step aims at improving the robustness of the initialisation step with respect to noise and CIB. We would like to highlight that the filtered data are only used during the initialisation; the final estimate of the dust estimation is performed on the raw data. Filtering consists of disentangling between the dust emission and the nuisance components, this can be achieved thanks to i) the Gaussianity of the CIB and noise components and ii) the sparsity of the dust emission in the wavelet domain.

In actuality, instrumental noise is faintly correlated across the sky as it follows the instrument's scanning strategy and the CIB anisotropies are also faintly spatially correlated as they trace galaxy clustering. However as these correlations are so faint we choose, as do \citet{gnilc}, to work under the standard assumption that both the CIB and instrumental noise are Gaussian.

The GNILC filtering requires an estimation of the nuisance terms from which to calculate the nuisance covariance matrix from. For our implementation of GNILC the CIB, instrumental noise and CMB estimates are in fact the actual CIB, instrumental noise and CMB contributions to the total flux. As {\tt{premise}} is intended for use after subtraction of the CMB we too use the same CIB and instrumental noise estimates as our GNILC-like implementation but this differs to the actual CIB and instrumental noise present in our total flux maps (see section \ref{sec:simCIB}).

One of the major advantages of GNILC is the ability to exploit the statistical properties of CIB and noise to disentangle between these components and the dust emission. Hence, the filtering stage will be based on a similar data processing with three major differences: 
 
\begin{itemize}
\item As described in Appendix \ref{sec:apB}, GNILC identifies the dust emission subspace by thresholding the eigenvalues of the data covariance matrix in the needlet domain after nuisance-whitening. For that purpose, it relies on a dimensionality selection criterion called AIC (Akaike Information Criterion), which turns out to be slightly conservative in practice. Consequently, the AIC tends to provide an over-smoothed estimation of the filtered data. In the proposed approach, the threshold is based on the actual statistics of the eigenvalues of the nuisance covariance. Since both the CIB and noise follow Gaussian distributions with known covariance matrices, their eigenvalues after whitening follow a Marcenko-Pastur distribution \citep{Mehta}. The dust emission subspace is estimated by identifying the eigenvalues which exceed a particular threshold, so chosen to ensure that the false detection probability is lower than $ 10^ {-3}$ according the Marcenko-Pastur distribution. This procedure is similar in spirit to the $3 \sigma$ thresholding used in standard Gaussian statistics. \\

\item GNILC further relies on a Gaussian smoothing of the data covariance matrices, which dramatically increases the computational cost of the filtering step. Our filtering is performed on overlapping patches or super-pixels with an overlapping ratio of $0.5$. This results in a decrease in the computational time required for this step without significantly altering the filtering quality.\\ 

\item The proposed GNILC-based filtering only exploits the statistical properties of CIB and noise; it does not profit from the sparsity of dust emission in the wavelet domain. Additionally, since it relies on a spatial smoothing of the data, it yields a smooth estimate (${\tilde{\bf{X}}}^{\rm{filt}}$) of the dust emission with lower spatial resolution. We attempt to mitigate the effects of smoothing by exploiting the sparsity of the dust emission in the wavelet domain. Since neither CIB nor noise have a sparse representation in the wavelet domain, we propose to enhance the filtering stage by updating the above filtered dust with sparse deviations in the wavelet domain. These sparse deviations can only originate from dust emission. Formally, each wavelet band of the estimated dust emission will be obtained as $\tilde{\bf W}_j^{\rm{dust}}  = \tilde{\bf W}_j^{\rm{filt}} + {\bf \Delta}_j$. The sparse deviation term ${\bf \Delta}_j$ is computed at each wavelet scale by solving a minimisation that is analogous to a sparse denoising problem \citep{starck2010sparse}:

\begin{eqnarray}
\min_{{\bf \Delta}_j} \lambda_j \|{\bf \Delta}_j \|_{2,1} + \frac{1}{2}\left\| {\bf W}_j - \tilde{\bf W}_j^{\rm{filt}} - {\bf \Delta}_j \right\|_F^2,
\end{eqnarray}
where ${\bf W}_j$ is the raw data and $\|\, . \,\|_F^2$ is the Frobenius norm. The L2,1-norm is defined for some matrix $\bf Y$ as:
\begin{eqnarray}
 \|{\bf Y}\|_{2,1} = \sum_k \sqrt{\sum_i{Y_{\nu_i}[k]^2}}. 
 \end{eqnarray}
 
Unlike the L1-norm, this regularisation term allows us to enforce the column-wise sparsity of the data, which is adapted to capture signals that have sparse distributions in space and are likely to be present in all channels. The problem admits a solution with a closed-form expression \citep{kowalski2009sparse} so that at each pixel:
 
\begin{equation}
\Delta_j[k] =  \left\{\begin{array}{l}
({\bf W}_j[k] - \tilde{\bf W}^{\rm{filt}}_j[k]) \times\left( 1 - \frac{ \lambda_j }{\| {\bf W}_j[k] - \tilde{\bf W}^{\rm{filt}}_j[k] \|_2^2}\right). \\
0 \mbox{ otherwise.}
\end{array}
\right.
\end{equation}

The solution is therefore nonzero whenever the $\chi^2$ of the residual ${\bf W}_j[k] - \tilde{\bf W}^{\rm{filt}}_j[k]$ is lower than the value $\lambda_j$ and equal to a pruned version of the residual otherwise. Since $\lambda_j$ plays the role of a threshold on the residual it can be chosen based on the statistical significance of the measured $\chi^2$. At each scale $\lambda_j$ is fixed to 2.7, which corresponds to a p-value of 10 per cent.

\end{itemize}

Combining both a GNILC-based filtering and wavelet-based filtering enables us to account for the statistical properties of the CIB and noise components as well as the sparsity of the dust emission in the wavelet domain.

\paragraph*{Quadtree-based fitting in the wavelet domain}

\begin{figure}
\centering
\subfloat{\includegraphics[width=0.99\linewidth]{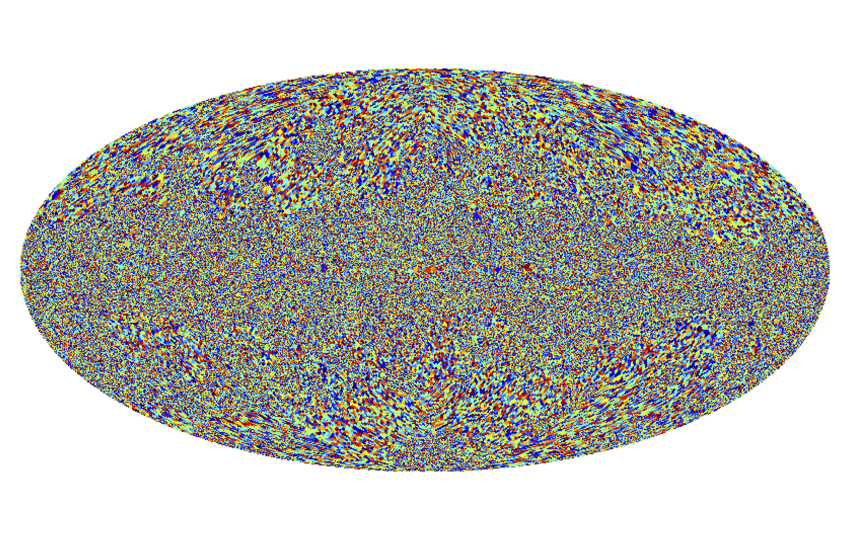}}
\caption{All-sky map of the different patches chosen by the quadtree.}
\label{fig:patches}
\end{figure}

Our fast, initial parameter estimation technique makes use of a quadtree. A quadtree recursively divides the given data into quarters until a particular criterion is no longer achieved within the data patch. The quadtree used in this method recursively divides a square of data until the number of data points (in this case, reduced $\chi^{2}$ values) with a value greater than 2 within the patch are less than 10 per cent of the total number of data points. Fig.~\ref{fig:patches} shows the patches selected by the quadtree over the full sky. The patches sizes are visibly smaller within and near to the Galactic plane where the total flux across neighbouring pixels is less consistent than at high latitudes. Very close to the Galactic centre several patch sizes larger than their neighbouring patches can be seen: these areas contain so many masked pixels (due to point sources) that the quadtree is forbidden from dividing the region into smaller areas. 

The {\texttt{premise}} fitting process works on each of the twelve 2048X2048, {\texttt{HEALPix}} 2D faces which make up the full sky as follows:
\begin{enumerate}[label*=\arabic*.]
\item The noise covariance matrix (${\bf{R}}^{n}$) of the full data face is calculated within the wavelet domain:
\begin{eqnarray}
{\bf{R}}_{j}^{n} = \frac{1}{2048^2} \left( {\bf{W}}_{j}^{n} \;  ({\bf{W}}_{j}^{n})^{T} \right),
\end{eqnarray} 
where 
\begin{eqnarray}
\nonumber
 {\bf{W}}_{j}^{n} =  {\bf{W}}_{j}^{\rm{noise}}  + {\bf{W}}_{j}^{\rm{CIB}} - {\bf{W}}_{j}^{\rm{CIB \, offset}} 
 \end{eqnarray} 
\item The region is split into patches of 128X128
\item Each patch is treated as a super-pixel: the mean flux density for each frequency represents the whole patch at that frequency	
\item A MBB fit to each super-pixel yields the parameters required to form a thermal dust estimate for each frequency at that super-pixel.
\item The data-model residual is calculated per original pixel:
$$
\rm{Residual}[k] = \rm{Total \, Flux}[k] -  \rm{Estimated \, thermal \, dust \, flux}[k] 
$$
\item This residual matrix (which has dimensions of number of frequencies by number of pixels) is transformed into the wavelet domain (${\bf{W}}_{j}^{r}$). 
\item The reduced $\chi^{2}$ at each of the wavelet scales is calculated for each pixel:
\begin{eqnarray}
\chi^{2}_{\rm{red}} =  ({\bf{W}}_{j}^{r})^{T}  ({\bf{R}}_{j}^{n})^{-1} \, {\bf{W}}_{j}^{r} 
\end{eqnarray}
As there is only one degree of freedom the reduced $\chi^{2}$ is just the $\chi^{2}$.
\item The reduced $\chi^{2}$ is fed back into the quadtree to split the data into patches where $T$ and $\beta$ are constant enough to be well characterised by a MBB with a single value of $T$ and $\beta$ across all pixels within the patch. The final patches chosen are the quadtree determined super-pixels.  
\item The fit is re-run fit on the quadtree chosen super-pixels and initial maps of $T$ and $\beta$ estimates are obtained.
\end{enumerate}

\paragraph*{Tuning the quadtree parameters}
\begin{itemize}[label={--}]
\item{Wavelet scales:} For steps 1. and 5. and 7. the number of wavelet scales used is four. For step 8. the higher wavelet scales are used for larger patch sizes so when deciding whether or not to split a 64X64 patch the quadtree uses the reduced $\chi^{2}$ values from the third wavelet scale, for a 16X16 patch the first wavelet scale is used. \\

\item{Patch area:} A lower limit of 8 is enforced to ensure that the quadtree cannot split the data into patches smaller than 8X8 pixels. The quadtree is also prohibited from splitting a patch into quarters if one or more of those quarters contains only masked data.
\end{itemize}

\quad 

Running the MBB fit on super-pixels, as opposed to the true pixels, reduces the computational time significantly.

\subsubsection{Parameter refinement} \label{parref}

In this section, we detail how the MBB parameters are estimated from the raw data by solving the penalised least-square estimation problem in Equation~\ref{eq:nl_sparseLS}. The main difficulty lies in the non-differentiability of the L1-norm. Consequently, minimising the cost function (Eq.~\ref{eq:cost_function}) cannot be achieved using standard gradient descent methods. The recently introduced proximal algorithms \citep{parikh2014proximal} can provide the perfect framework to design an effective minimiser for the problem. We will make use of the forward-backward splitting algorithm: an iterative projected gradient descent algorithm (see \citet{parikh2014proximal} and references therein) which has been specifically designed to solve penalised least-square problems with linear models. As a solution to the non-linear problem Eq.~\ref{eq:cost_function}, the proposed approach can be considered as an extension of the projected gradient descent techniques introduced in \citep{teschke09}.

The parameter optimisation step of {\texttt{premise}} builds upon an iterative estimation procedure, composing of the following two stages:
\begin{itemize}
\item{\it Gradient descent: } this steps consists in performing a single gradient descent step of the data fidelity term with respect to $\beta$ and $T$:
\begin{multline}
\label{eq:cost_function}
\beta^{(\frac{1}{2})}[k]  =  \beta^{(t)}[k] + \\ \alpha g_{\beta}^{T}[k] {\bf \mathcal{M}}[k]  \left(X[k] -  {\bf \mathcal{M}}[k]  X[k]^{\mbox{dust}}(\beta^{(t)}[k],T^{(t)}[k])  \right), \nonumber
\end{multline}
\begin{multline}
T^{(\frac{1}{2})}[k]  =  T^{(t)}[k] + \\ \alpha g_{T}^{T}[k] {\bf \mathcal{M}}[k]  \left(X[k] - {\bf \mathcal{M}}[k]  X[k]^{\mbox{dust}}(\beta^{(t)}[k],T^{(t)}[k])  \right), 
\end{multline}
where  $\alpha$ is the gradient path length and $g_{Y}$[k] is a vector whose $i$-th is equal to the derivative of the MBB model:
$$
\frac{\partial  X_{\nu_i}^{\mbox{dust}}[k]}{\partial Y[k]}(\beta[k]^{(t)},T[k]^{(t)})  ,
$$
where $Y$  indicates what the derivative is in respect to (in this case $\beta$ or $T$). These quantities are computed analytically from the expression of the MBB model.\\

\item{\it Projection step: } In this step, the estimated $\beta$ and $T$ maps are thresholded in the wavelet domain:
\begin{eqnarray}
\beta^{(t + 1)}[k] & = & \mathcal{S}_{\alpha \lambda_{\beta}}\left( \beta^{(\frac{1}{2})} {\bf \Phi} \right) {\bf \Phi}^{-1}, \\
T^{(t + 1)}[k] & = & \mathcal{S}_{\alpha \lambda_{T}}\left( T^{(\frac{1}{2})} {\bf \Phi} \right) {\bf \Phi}^{-1}.
\end{eqnarray}

\begin{center}
	\centering
	\vspace{0.25in}
	\begin{tabular}{|c|} \hline
		\begin{minipage}[hbt]{0.95\linewidth}
			\vspace{0.15in}
			
			\textsf{{\bf Initialise} $\beta^{(0)}$ and $T^{(0)}$ with the quad-tree based fitting technique,}\\
			
			\textsf{\bf At each iteration $t$.}\\
			
			\hspace{0.2in} \textsf{1 - Gradient descent step for each pixel} \\ \\
				\hspace{0.4in}  $\beta^{(\frac{1}{2})}[k]  =  \beta^{(t)}[k] +  \\  \alpha g_{\beta}^T[k] {\bf \mathcal{M}}[k]\left({X}[k] - {\bf \mathcal{M}}_{\nu_i}[k]{X}^{\mbox{dust}}[k](\beta^{(t)}[k],T^{(t)}[k])  \right) $\\
				\hspace{0.4in}  $T^{(\frac{1}{2})}[k]  =  T^{(t)}[k] + \\ \alpha g_{T}^T[k] {\bf \mathcal{M}}[k] \left({ X}[k] - {\bf \mathcal{M}}_{\nu_i}[k]{X}^{\mbox{dust}}[k](\beta^{(t)}[k],T^{(t)}[k])  \right) $\\
			
			\hspace{0.2in} \textsf{2 - Thresholding step} \\ \\
				\hspace{0.4in}  $\beta^{(t + 1)}[k]  =  \mathcal{S}_{\alpha \lambda_{\beta}}\left( \beta^{(\frac{1}{2})} {\bf \Phi} \right) {\bf \Phi}^{-1} $\\
				\hspace{0.4in}  $T^{(t + 1)}[k]  =  \mathcal{S}_{\alpha \lambda_{T}}\left( T^{(\frac{1}{2})} {\bf \Phi} \right) {\bf \Phi}^{-1}$ \\

			\textsf{\textbf{Stop} when a given criterion is valid.}
			
			\vspace{0.15in}
		\end{minipage}
		\\\hline
	\end{tabular}
	\vspace{0.25in}
\end{center}

\paragraph*{Tuning the refinement parameters}
\begin{itemize}
\item{Gradient path length:} The algorithm converges to the stationary point of the problem in Eq. \ref{eq:cost_function} provided that $\alpha \leq  \min_k 1/||H[k]||_2$, where $H$ is the Hessian matrix of the data fidelity term with respect to $\beta$ and $T$. The spectral norm, $\| H[k]\|_2$, is the largest singular value of the Hessian matrix. The proposed iterative thresholded gradient descent technique builds upon a local linear approximation of the data fidelity term about the current estimate at iteration $t$. The same linear approximation yields $\alpha \leq  \min_k 1/(g_{\beta}^T[k] g_{\beta}[k] +g_{T}^T[k] g_{T}[k])$, which is computed analytically at each iteration.\\

\item{Thresholds:} A careful choice of the thresholds $\lambda_b$ and $\lambda_t$ are essential since these thresholds tune the trade-off between the data fidelity term and the sparse penalisation. For that purpose, we make use of a very effective, heuristic approach to choose these parameters automatically. The goal of the thresholding stage is to save the wavelet coefficients with significant amplitude while removing noise or non-sparse elements. In this context, the standard deviation of the non-sparse elements can be computed using the MAD (Median Absolute Deviation) \citep{starck2010sparse}; the thresholds are then chosen as $\lambda_b = 2\, \sigma_{\mbox{MAD}(\beta^{(\frac{1}{2})})}$ and $\lambda_T = 2 \, \sigma_{\mbox{MAD}(T^{(\frac{1}{2})})}$. In practice, the thresholds are computed independently at each of the five wavelet scales. \\

\item{Stopping criterion:}	The proposed iterative algorithm stops when the relative variation of the estimates between two consecutive iterations is lower than a given level: 
\begin{eqnarray}
 \max_{\beta,T} \left(  \frac{\|\beta^{(t+1)}[k] - \beta^{(t)}[k] \|_{\ell_2}}{\|\beta^{(t)}[k] \|_{\ell_2}},   \frac{\|T^{(t+1)}[k] - T^{(t)}[k] \|_{\ell_2}}{\|T^{(t)}[k] \|_{\ell_2}}   \right) \leq 10^{-6} \nonumber
\end{eqnarray}
\end{itemize}

\end{itemize}

\subsubsection{Optical depth at 353\,GHz}

Up to this point we have only dealt with two of the three MBB parameters: temperature and spectral index. Both the fast wavelet-based initialisation and the parameter refinement steps can be utilised to produce optical depth values just as they do for the temperature and spectral index. However, due to the degeneracies between the three parameters introduced through residual CIB and noise, {\texttt{premise}} produces less accurate optical depth estimates than desired. Both \citet{pr2} and \citet{gnilc} encounter these degeneracies and deal with them through smoothing; in \citet{pr2} the spectral index maps are smoothed to 30 arcmin while in \citet{gnilc} all three parameters are smoothed within the lowest signal-to-noise regions of the sky. 

{\texttt{premise}} is specialised to trace the spatial variations of the dust temperature and spectral index, to the detriment of the overall normalisation factor (the optical depth). At this point it is therefore useful to introduce additional information to break the parameter degeneracies and recover the normalisation factor: a thermal dust template. The total flux 857\,GHz map is known to have minimal CIB contamination and so provides us with the opportunity to recover the 353\,GHz optical depth at 5 arcmin using the {\texttt{premise}} estimates for temperature and spectral index: 
\begin{eqnarray}
\tau_{353} = \frac{x_{857} \times \rm{cc}}{B(T,  857\,\rm{GHz}) \times \left(\frac{\nu}{353 \, \rm{GHz}}\right)^{\beta} }.
\end{eqnarray}

As the total flux data include point sources, a masked and inpainted \citep{inpaint} version of the total flux 857\,GHz map is used to recover the optical depth at 353\,GHz. As fractional CIB contamination is present within the 857\;GHz total flux data, this contamination will propagate through to the {\texttt{premise}} optical depth estimate. In the following section we will determine if the use of the 857\,GHz total flux data is an acceptable way to recover optical depth information or if the CIB contamination present at 857\,GHz is still too high to produce reliable optical depth estimates.

\section{Results}
\label{sec:results}

\subsection{Regions}

\begin{figure}
	\centering
	\includegraphics[width=0.99\linewidth]{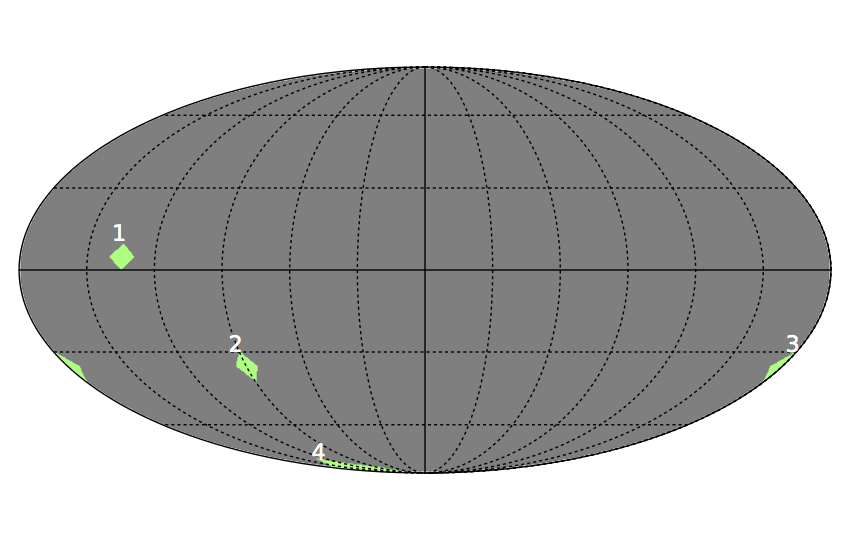}
	\caption{The location on the sphere of the four 256X256 pixel regions chosen for analysis.}
	\label{fig:globe}
\end{figure}

We compute {\texttt{premise}} on the full sky, however as GNILC is a computationally intensive algorithm we only run the GNILC-like methodology on four test regions and conduct our comparison within these four, 256X256 pixel regions. The regions comprise of a high signal-to-noise region in the Galactic plane (region 1), two medium signal-to-noise regions at intermediate latitudes - one at a central longitude (region 2) and one at zero longitude (region 3) and a low signal-to-noise polar region (region 4). The location of these regions within a {\texttt{HEALPix}} sphere is shown in Fig.~\ref{fig:globe}. 

\begin{figure*}
\centering
\subfloat{\includegraphics[width=0.4\linewidth]{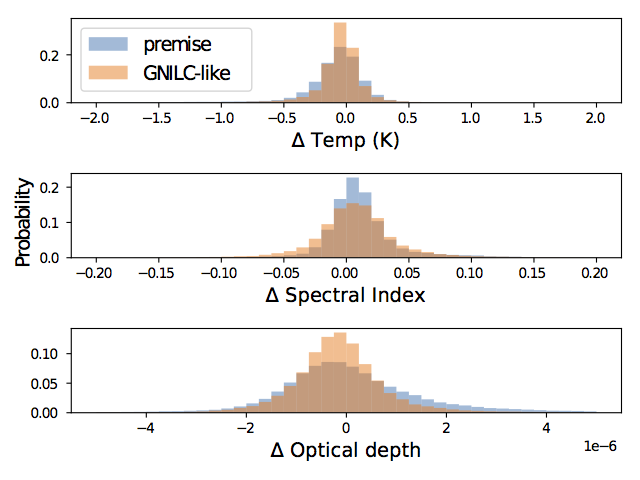}} 
\subfloat{\includegraphics[width=0.4\linewidth]{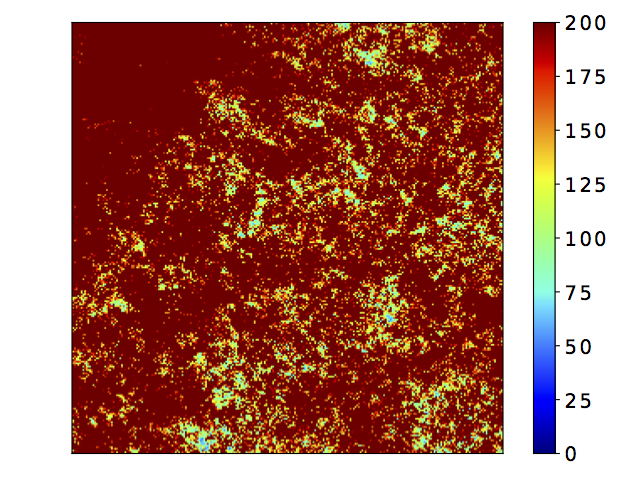}} \\
\subfloat{\includegraphics[width=0.4\linewidth]{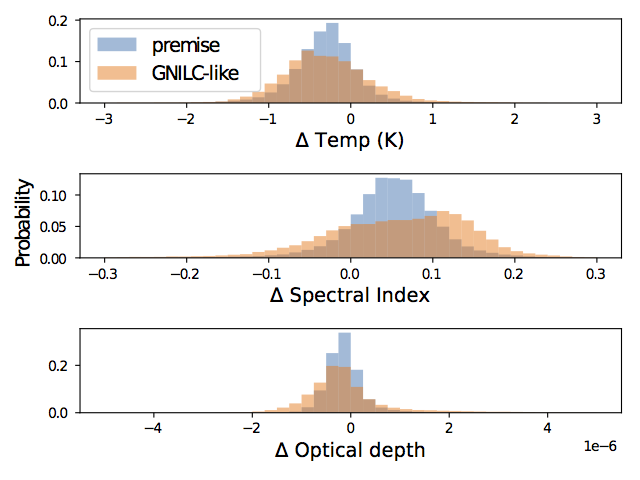}}
\subfloat{\includegraphics[width=0.4\linewidth]{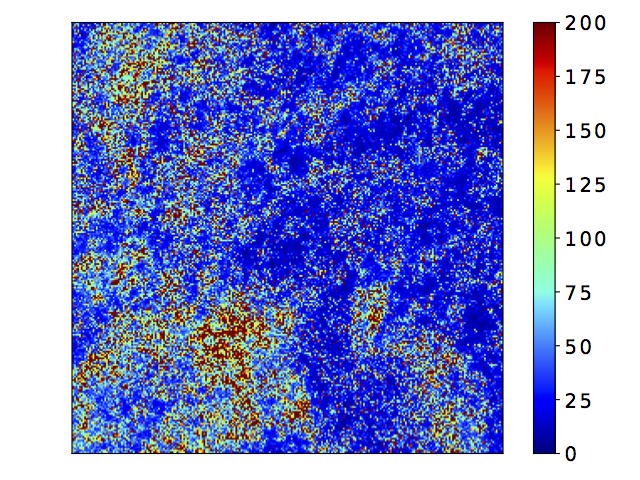}}\\
\subfloat{\includegraphics[width=0.4\linewidth]{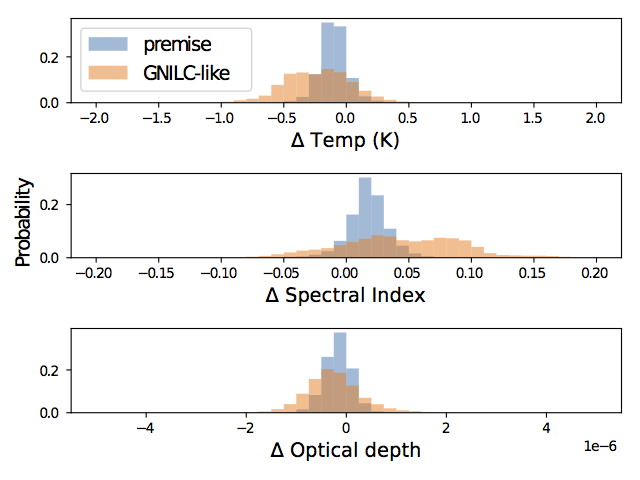}}
\subfloat{\includegraphics[width=0.4\linewidth]{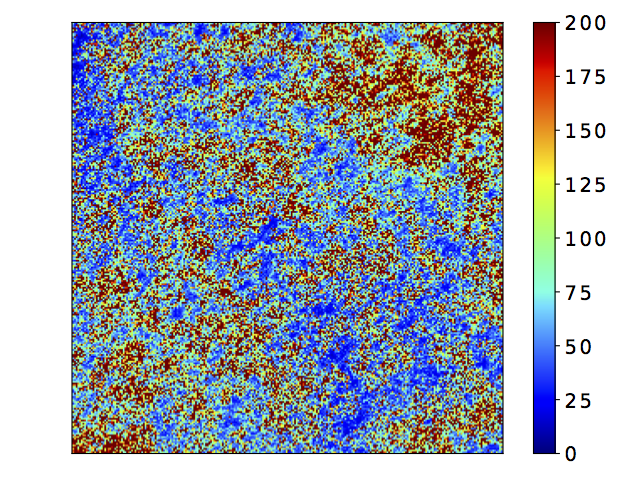}}\\
\subfloat{\includegraphics[width=0.4\linewidth]{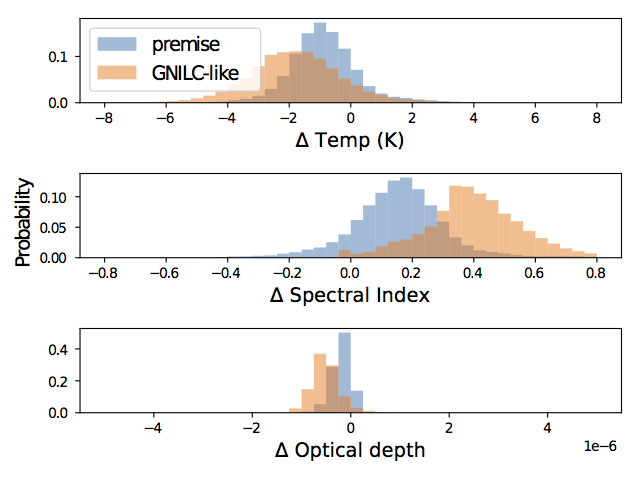}}
\subfloat{\includegraphics[width=0.4\linewidth]{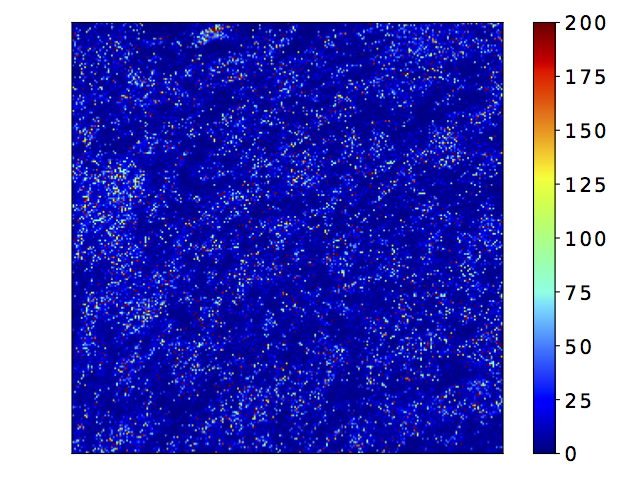}}\\
\caption{{\it{Left:}} Histograms of the differences between the calculated and true MBB parameters for the GNILC-like methodology and {\texttt{premise}}. {\it{Right:}} Signal-to-noise maps. Results for region 1 to 4 shown from top to bottom.}
\label{fig:histAll}
\end{figure*}

\subsection{Comparison of parameter estimation within Regions 1--4}

In the following section we compare the GNILC-like and {\texttt{premise}} fitted $\tau_{353}$, $T$ and $\beta$ values for Regions 1--4 to the true $\tau_{353}$, $T$ and $\beta$. Fig.~\ref{fig:histAll} shows histograms of the actual differences between the true parameter values and those derived from the GNILC-like methodology/{\texttt{premise}}. Signal-to-noise maps (the ratio of pure thermal dust emission to the combination of instrumental noise and CIB) of each region are shown in the right-hand column. For region 1, the region with the largest signal-to-noise ratio, the GNILC-like $\tau_{353}$ and $T$ parameters show smaller differences to the true parameter values than {\texttt{premise}}. The opposite being true for the spectral index estimates. The GNILC-like methodology has an advantage over {\texttt{premise}}, at high signal-to-noise, with regards to the estimation of tau as the {\texttt{premise}} quadtree specialises in detecting spectral deviations between the MMB model and the data as opposed to selecting super-pixels sizes to best estimate the normalisation factor.  

\begin{figure*}
\centering
\subfloat[][]{\includegraphics[width=0.5\linewidth]{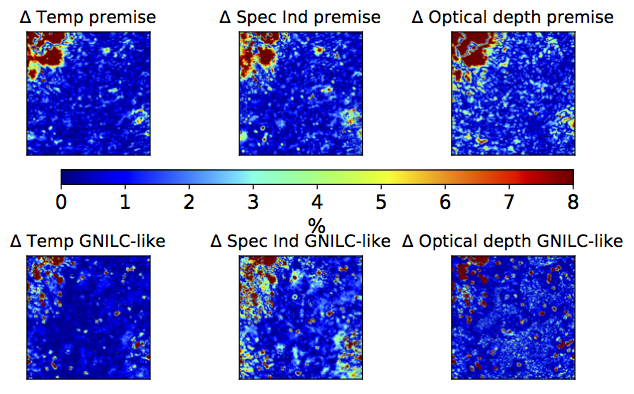}\label{a}}
\subfloat[][]{\includegraphics[width=0.5\linewidth]{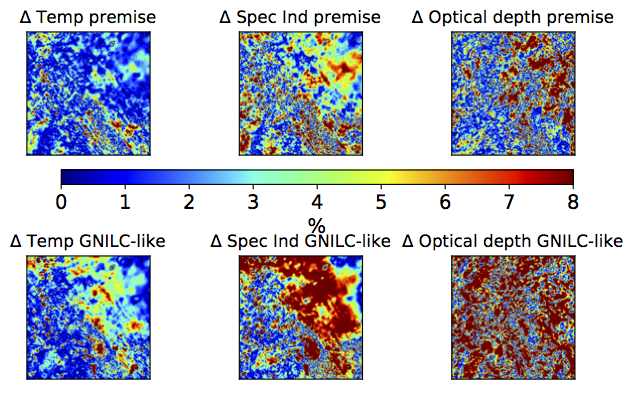}\label{b}}\,
\subfloat[][]{\includegraphics[width=0.5\linewidth]{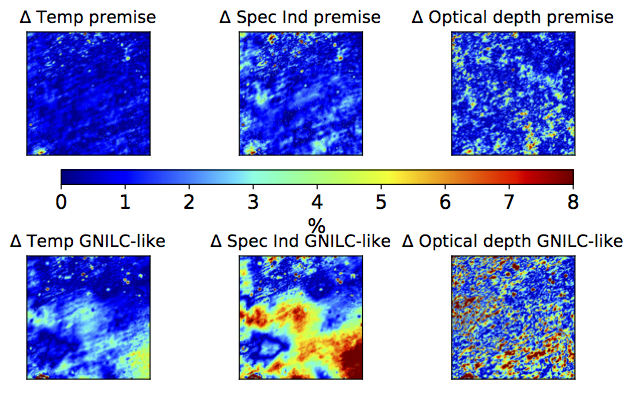}\label{c}}
\subfloat[][]{\includegraphics[width=0.5\linewidth]{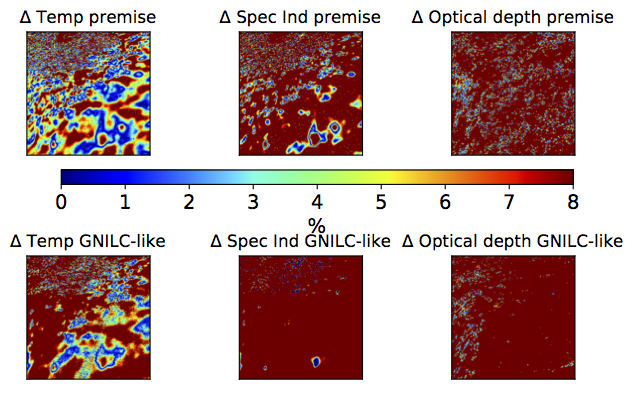}\label{d}}
\caption{Maps of absolute percentage differences between the calculated and true MBB parameters for the GNILC-like methodology and {\texttt{premise}} for each pixel. Results for region 1 \protect\subref{a},  2 \protect\subref{b}, 3 \protect\subref{c} and 4 \protect\subref{d}.}
\label{fig:mapsAll}
\end{figure*}

\begin{figure}
	\centering
	\includegraphics[width=0.99\linewidth]{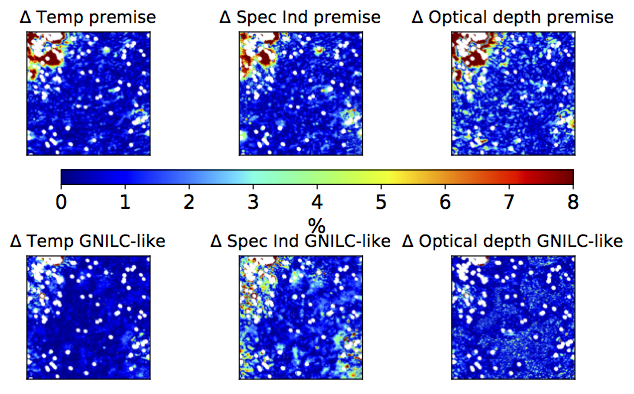}
	\caption{Region 1 map of absolute percentage differences between model and true temperature/spectral index for the GNILC-like methodology and {\texttt{premise}}. Point source masks applied}
	\label{fig:pointMap}
\end{figure}

Outside of the Galactic plane, however, smaller differences between the true and fitted parameters can be seen for {{\texttt{premise}}. This can be seen most clearly in regions 2 and 4, were the thermal dust signal-to-noise ratio is lowest. As the signal-to-noise drops the GNILC-like algorithm increases the degree of smoothing and their parameter estimates deviate further from the true values. Therefore, from the perspective of parameter estimation across a diverse range of signal-to-noise ratios, {\texttt{premise}} provides the more robust set of parameter values for the full sky.  

Fig.~\ref{fig:mapsAll} shows maps of the absolute percentage differences between the true parameter values and those derived from the GNILC-like methodology/{\texttt{premise}}. For region 1, the pixels with the poorest parameter estimates (for both methods) are those surrounded by clusters of extragalactic points sources. Fig.~\ref{fig:pointMap} is Fig.~\ref{fig:mapsAll} a) with the total point source mask (the individual frequency mask maps multiplied together) applied. The GNILC-like methodology uses inpainting to determine $\beta$ and $T$ within masked regions while {\texttt{premise}} just averages over larger super-pixel sizes than desirable for those regions where large portions of the data are masked out. As a result of this, the {\texttt{premise}} parameter estimates suffer most within regions of high point-source contamination, like the Galactic plane. 

Within regions 2, 3 and 4, Fig.~\ref{fig:mapsAll} reenforces the fact that the GNILC-like estimations suffer from over-smoothing as whole patches of the parameter space are misestimated. Region 4 is plotted using the same colour-scale as the other regions to highlight how poorly both methods fair (when compared to their optimal performances) within the lowest signal-to-noise regions of the full sky.

\begin{figure*}
\centering
\subfloat{\includegraphics[width=0.65\linewidth]{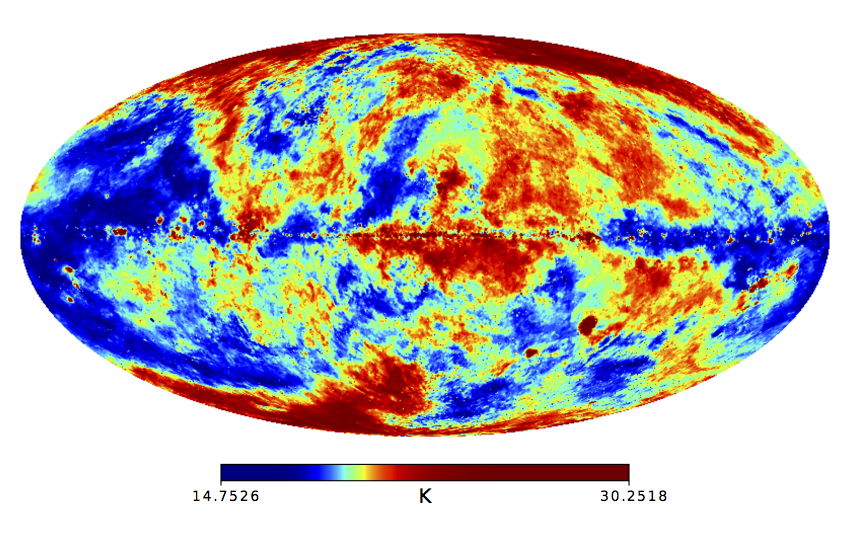}}\,
\subfloat{\includegraphics[width=0.65\linewidth]{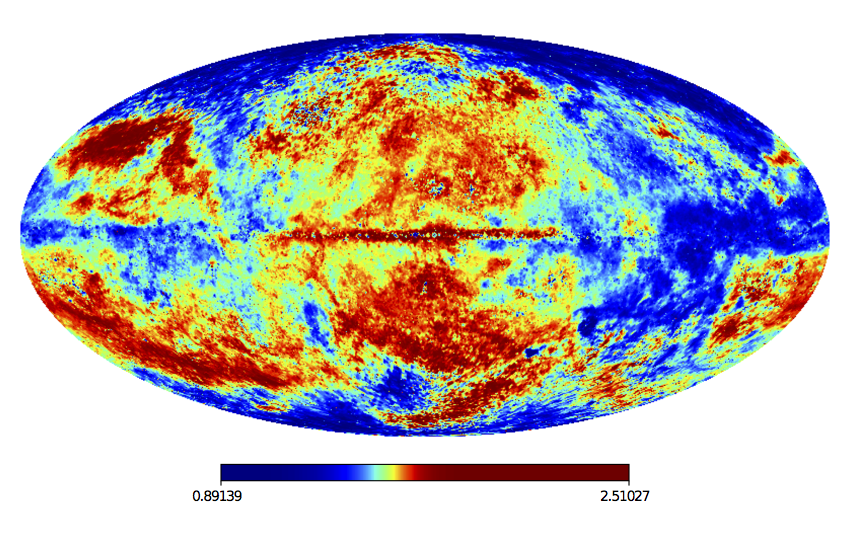}}\,
\subfloat{\includegraphics[width=0.65\linewidth]{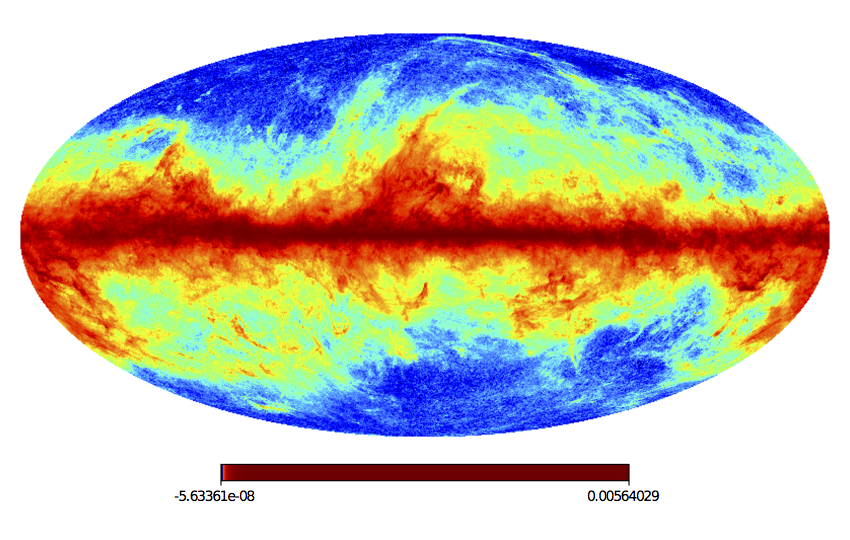}}\,
\caption{All-sky maps of thermal dust temperature ({\it{top}}), spectral index ({\it{middle}}) and optical depth at 353\,GHz  ({\it{bottom}}) produced by {\texttt{premise}}. The colour scale uses histogram equalisation.}
\label{fig:paramsmaps}
\end{figure*}

\begin{figure*}
\centering
\subfloat{\includegraphics[width=0.65\linewidth]{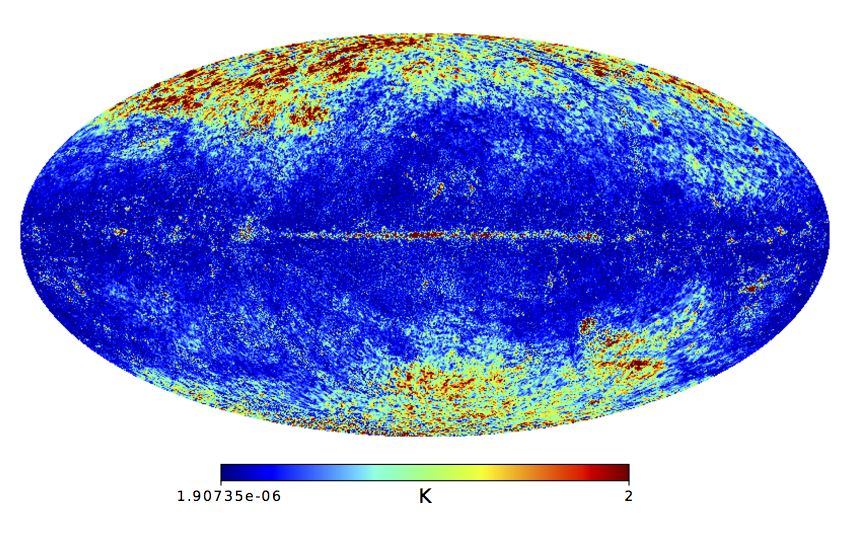}}\,
\subfloat{\includegraphics[width=0.65\linewidth]{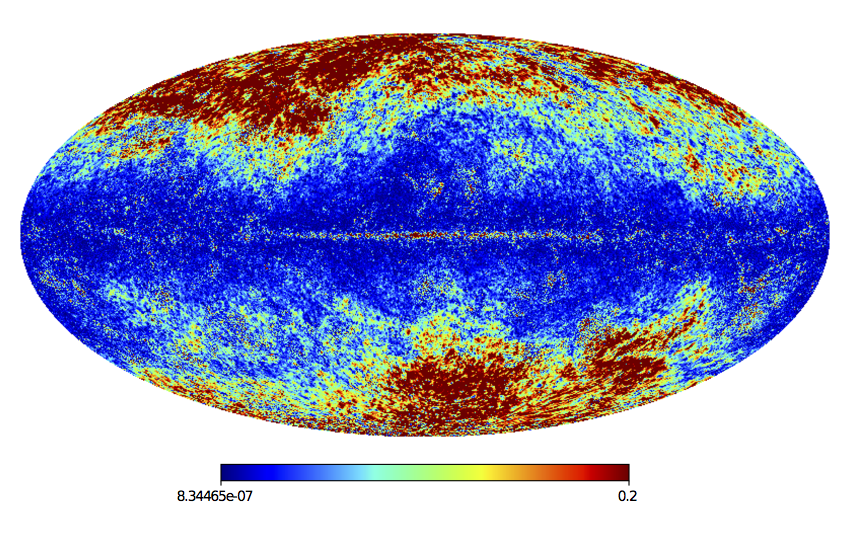}}\,
\subfloat{\includegraphics[width=0.65\linewidth]{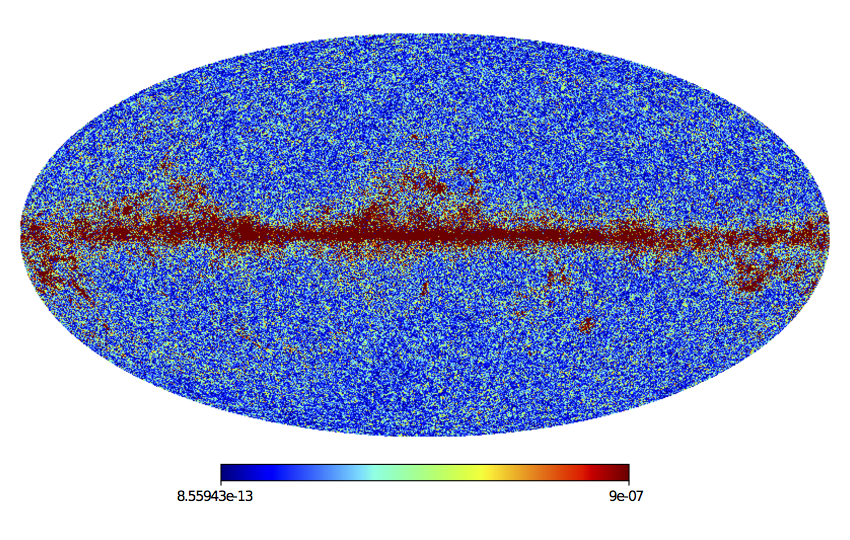}}\,
\caption{All-sky maps of the absolute differences between true and {\texttt{premise}} calculated values for thermal dust temperature ({\it{top}}), spectral index ({\it{middle}}) and optical depth at 353\,GHz  ({\it{bottom}}).}
\label{fig:paramsmapsDiff}
\end{figure*}

\subsection{Full sky {\texttt{premise}}}

\subsubsection{Parameters}

The all-sky {\texttt{premise}} thermal dust temperature, spectral index and optical depth at 353\,GHz are shown in Fig.~\ref{fig:paramsmaps}. Fig.~\ref{fig:paramsmapsDiff} displays the absolute differences between true and {\texttt{premise}} calculated parameters. From the difference maps it is clear that the estimation of both $\beta$ and $T$ suffer within the Galactic plane and at high latitudes. The result at high latitudes is unsurprising as the thermal dust signal-to-noise ratio is at its lowest within these regions. Yet {\texttt{premise}} does not perform optimally within the Galactic plane, despite the high signal-to-noise. This is due to the large number of point sources present. The increase in masked data towards the centre of the Galactic plane results in fewer pixels of information from which to determine the thermal dust parameters. The worst estimations of $\tau_{353}$ appear within the Galactic plane. This is simply because the largest optical depth values occur within the Galactic plane and so the absolute differences between true and estimated values appear largest.  

\begin{figure}
\centering
\subfloat{\includegraphics[width=0.95\linewidth]{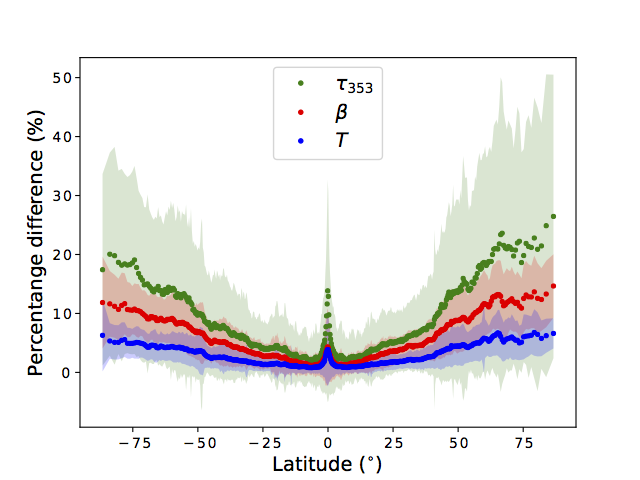}}\,
\caption{Mean absolute percentage difference between true and {\texttt{premise}} estimated parameter values as a function of Galactic latitude. Shaded area represent the standard deviation of the mean values.}
\label{fig:lats}
\end{figure}

Similarly to \citet{gnilc}, the success of {\texttt{premise}} is proportional to the thermal dust signal-to-noise ratio; this is highlighted in Fig.~\ref{fig:lats} which shows the mean absolute percentage difference between true and {\texttt{premise}} estimated values across latitudes (the data are binned in intervals of 512). The peak in percentage difference within the Galactic plane (latitude 0$^{\circ}$) is caused by a high fraction of the data being masked due to point sources. This peak can be seen to be larger for $\tau_{353}$ than for $\beta$ and $T$. This is because the {\texttt{premise}} calculated $\tau_{353}$ looses thermal dust information due to point source masking twice. The $\tau_{353}$ {\texttt{premise}} maps are made using the total 857\,GHz map after the masked data has been inpainted. The majority of this inpainting occurs within the Galactic plane and is an imperfect attempt to recover lost thermal dust information. This inpainted 857\,GHz map is then used to determine the dust optical depth at 353\,GHz using the {\texttt{premise}} determined temperature and spectral index values, which already suffer from a degradation in accuracy across the Galactic plane.

\begin{figure*}
\centering
\subfloat{\includegraphics[width=0.4\linewidth]{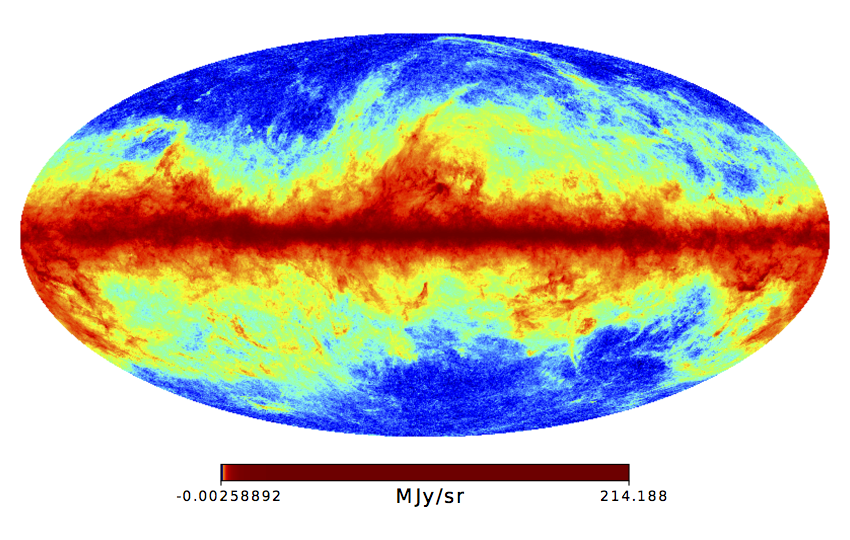}}
\subfloat{\includegraphics[width=0.4\linewidth]{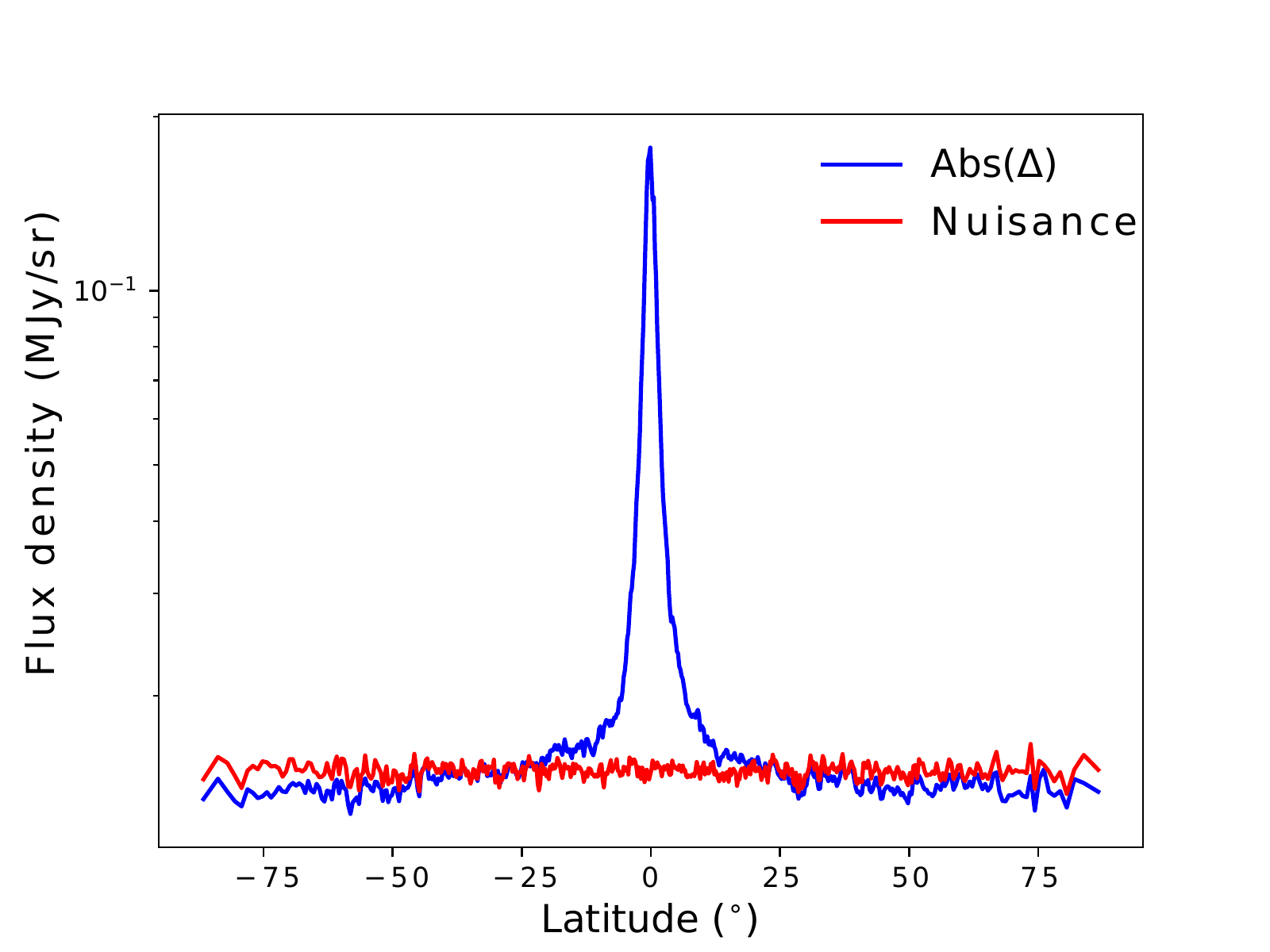}}\,
\subfloat{\includegraphics[width=0.4\linewidth]{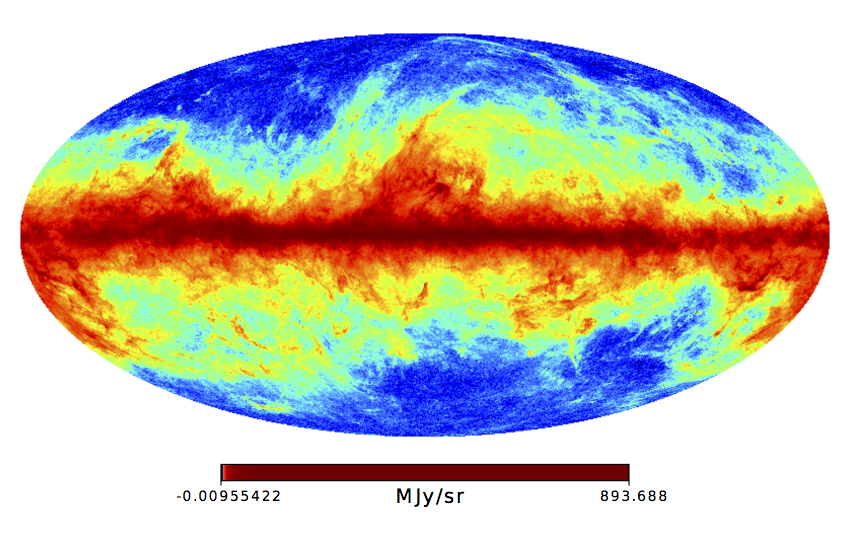}}
\subfloat{\includegraphics[width=0.4\linewidth]{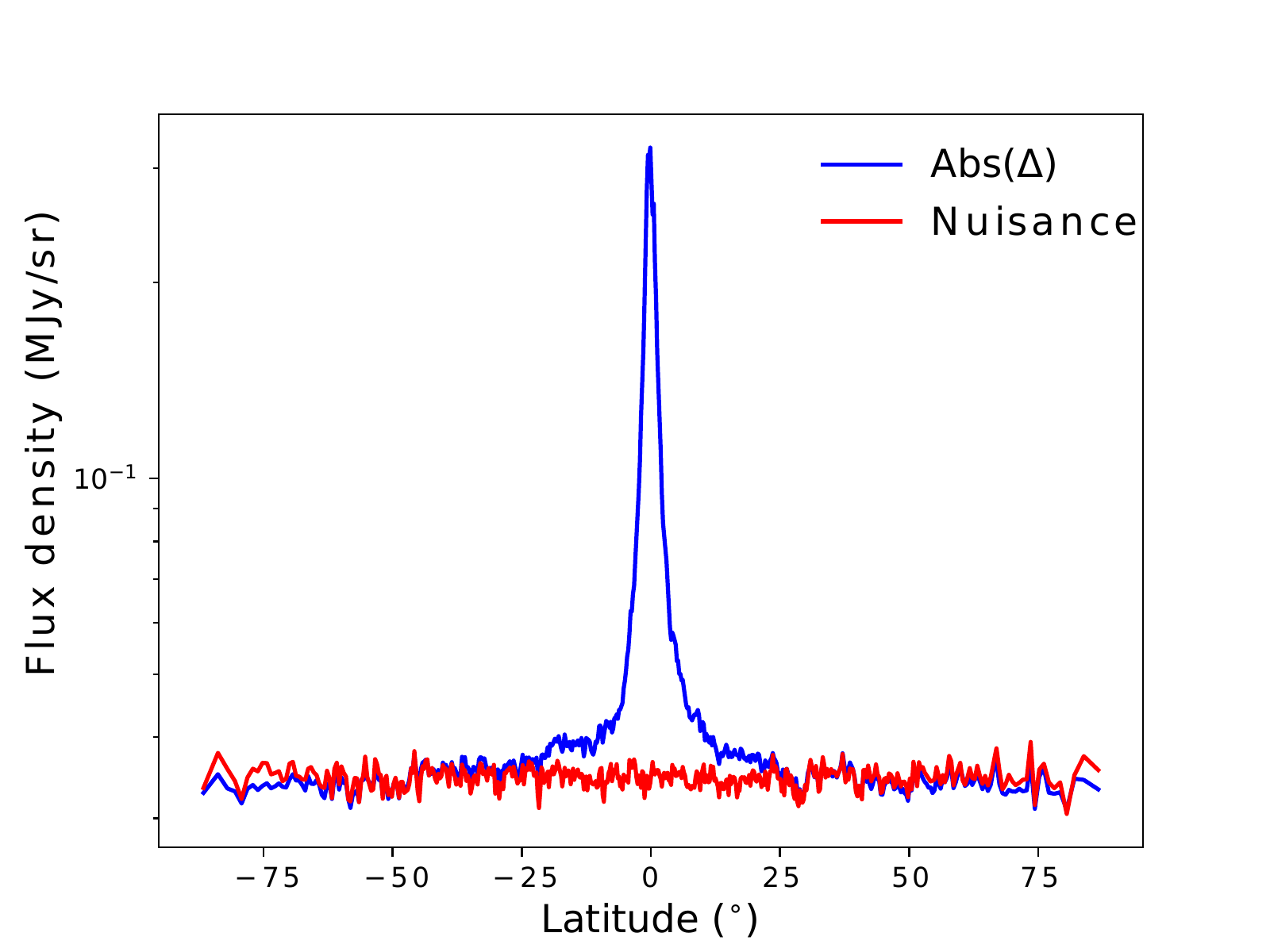}}\,
\subfloat{\includegraphics[width=0.4\linewidth]{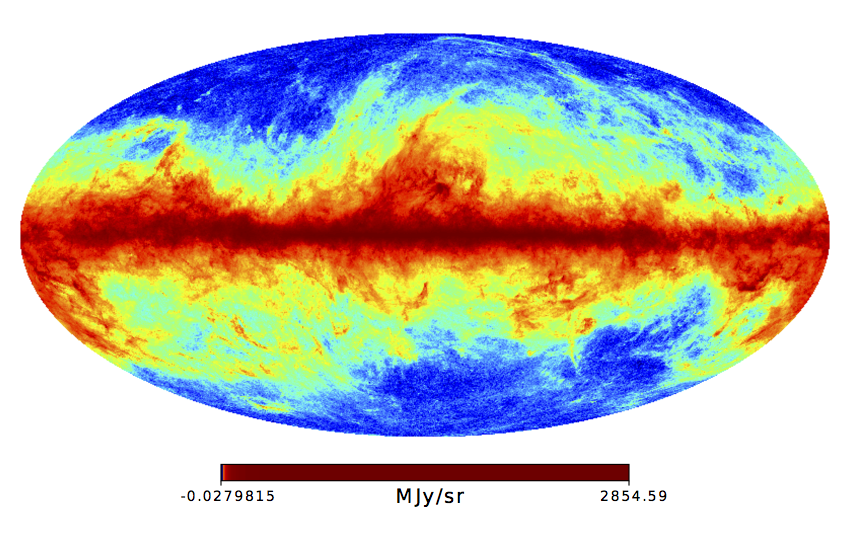}}
\subfloat{\includegraphics[width=0.4\linewidth]{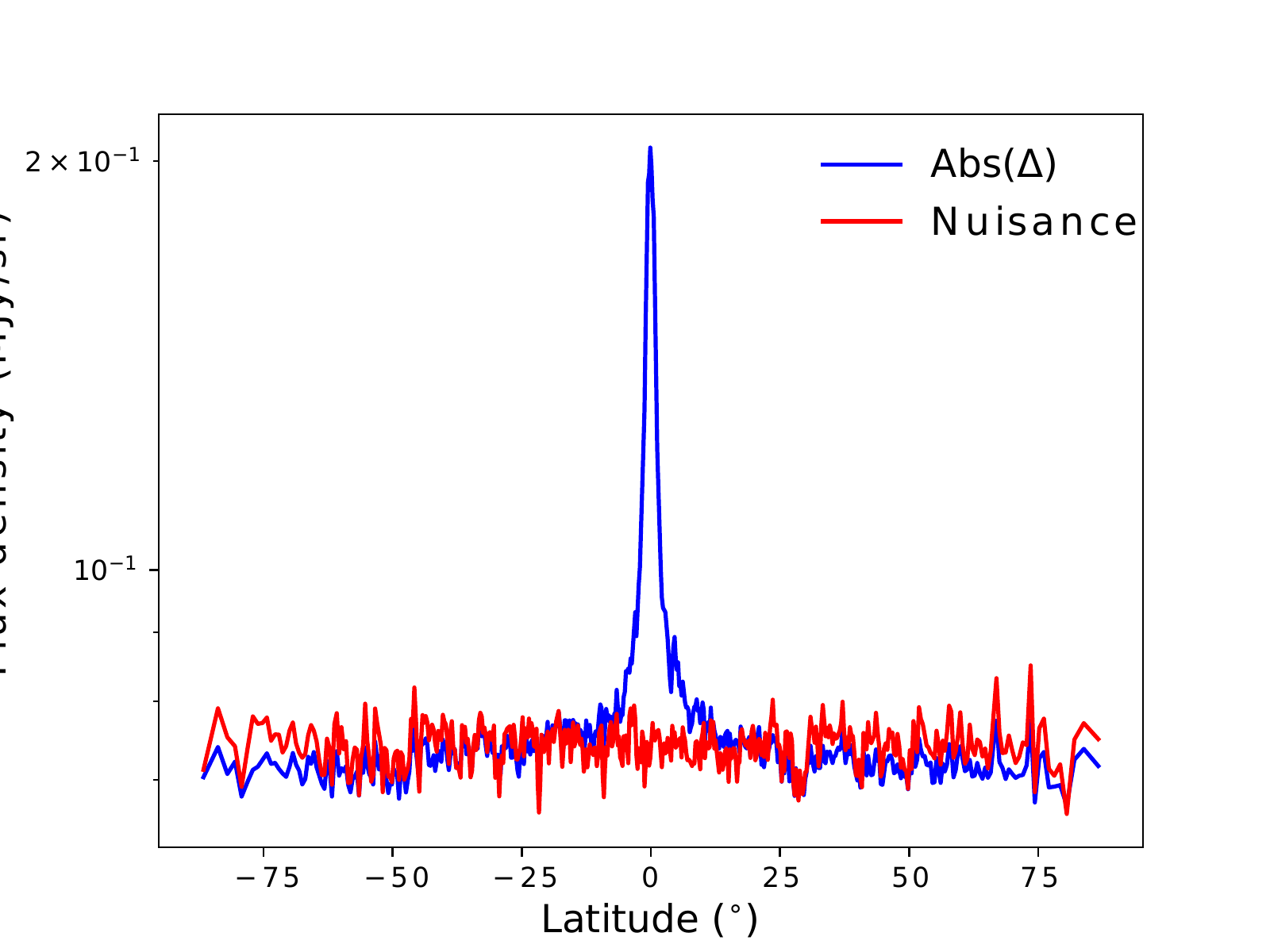}}\,
\subfloat{\includegraphics[width=0.4\linewidth]{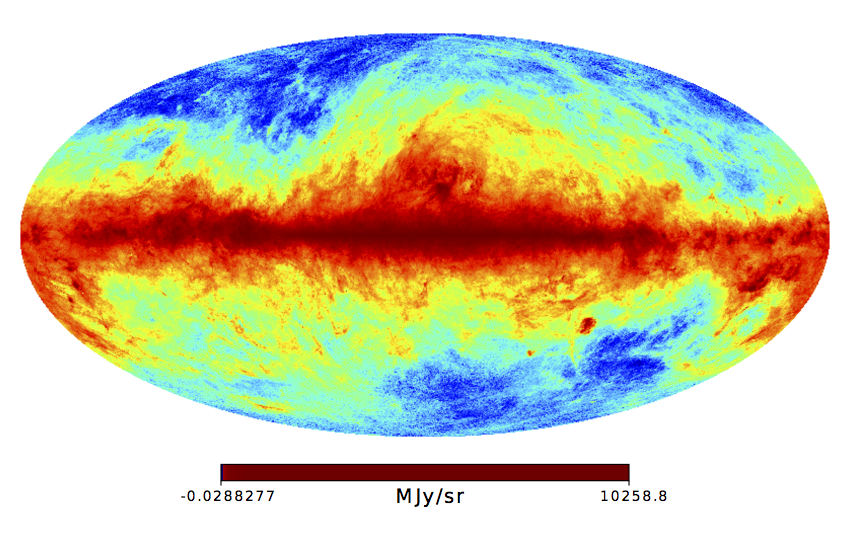}}
\subfloat{\includegraphics[width=0.4\linewidth]{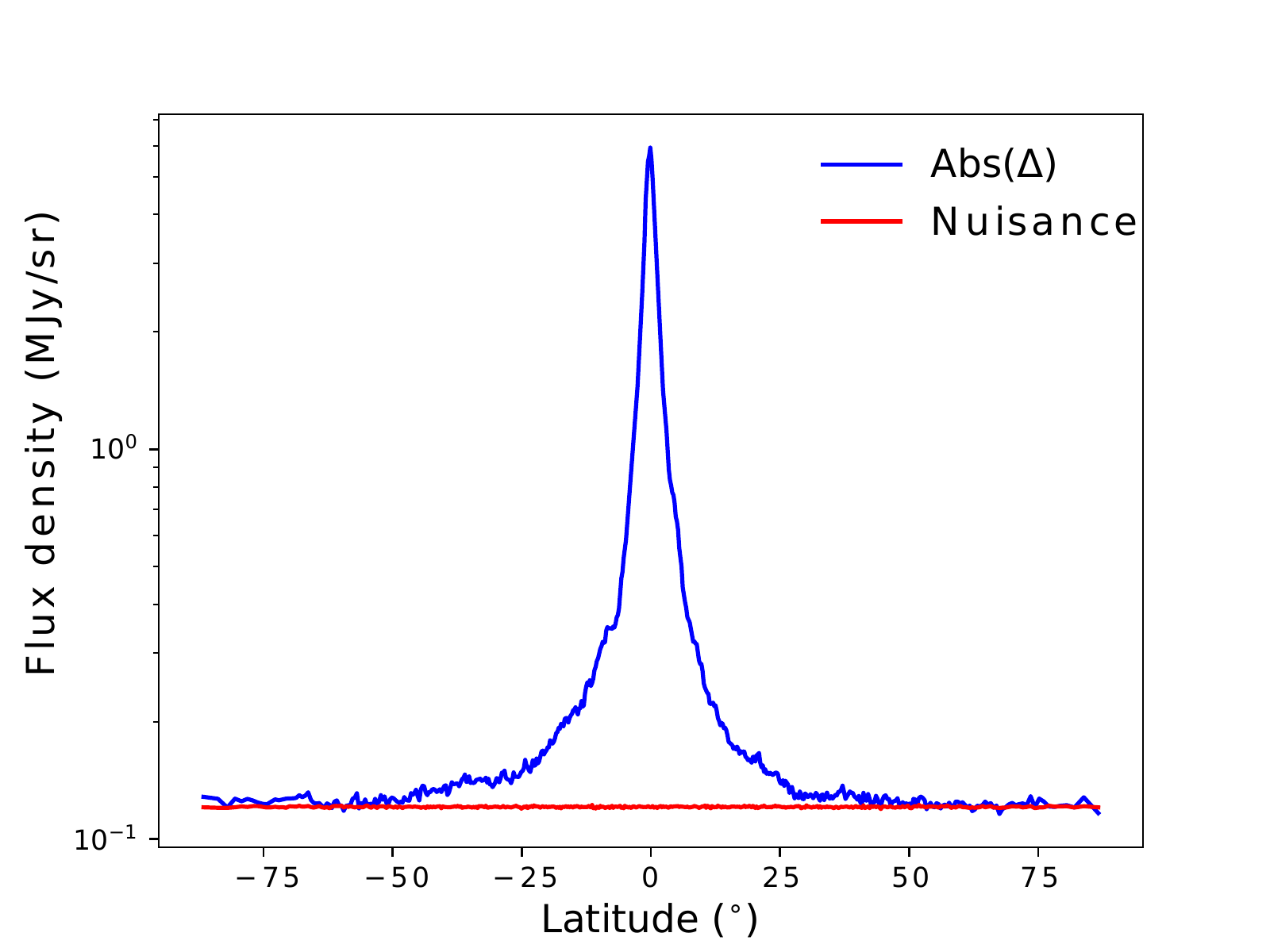}}\,
\caption{{\it{Left column:}} All-sky maps of thermal dust emission produced from the {\texttt{premise}} parameters. These maps have not been colour corrected and a histogram colour scale has been used. {\it{Right column:}} Plots of the absolute median differences between true and calculated thermal dust emission over latitude alongside the CIB anisotropies plus noise level (referred to as a `nuisance' term). Results at frequencies 353, 545, 857 and 3000\,GHz are shown from top to bottom.}
\label{fig:dustmaps}
\end{figure*}

\subsubsection{Thermal dust emission}

The three {\texttt{premise}} full-sky parameter maps can be reconstituted to form the {\texttt{premise}} thermal dust estimates at 353, 545, 857 and 3000\,GHz. We present these maps in the left-hand column of Fig.~\ref{fig:dustmaps}. The plots in the right-hand column show the median absolute flux density difference between the true and estimated dust maps when the data are binned at intervals of 512. We over-plot the nuisance flux term, CIB anisotropies (total CIB flux minus the constant CIB offset level) plus instrumental noise, in red. It can be seen that, for all frequencies, the largest discrepancies between true and estimated flux density occur within and approaching the Galactic plane. These discrepancies are far larger than the CIB and noise contamination simply because the thermal dust emission magnitude close to the Galactic plane is so high. On average the discrepancies are less than 6 per cent (for all frequencies): Table~\ref{tab:accstats} states the percentage difference between true and {\texttt{premise}} estimated thermal dust emission for 50 per cent (medium), 68.3 per cent (1$\sigma$), 95.4 per cent (2$\sigma$) and 99.7 per cent (3$\sigma$) of the full sky.

\begin{table}
 \caption{The percentage difference between true and {\texttt{premise}} estimated thermal dust parameters for various percentiles.}
 \label{tab:accstats}
 \begin{tabular}{l| c| c| c| c}
\hline
{\bf{Value}} &{\bf{ Median $\% \Delta$}}  & {\bf{1$ \sigma \, \% \Delta$}}  & {\bf{2$ \sigma \, \% \Delta$}}  & {\bf{3$ \sigma \, \% \Delta$}}  \\ 
Temperature & 1.7 & 2.8 & 8.0 & 16.5 \\
Spectral index & 3.4 & 5.7 & 15.4 & 25.6 \\
Optical depth (353\,GHz) & 3.7 & 7.2 & 31.2 & 77.0\\
353\,GHz emission & 5.1  & 10.1  & 39.9  & 94.9 \\
545\,GHz emission & 3.4 & 7.1 & 31.1 & 77.2 \\
857\,GHz emission & 2.5 & 5.3 & 25.3 & 66.4 \\
3000\,GHz emission & 4.5 & 7.7 & 29.8 & 83.4 \\
  \hline
 \end{tabular}
\end{table}

The 353 and 545\,GHz latitude plots in Fig.~\ref{fig:dustmaps} show absolute flux density differences at high latitudes at a slightly lower level than than the actual 353 and 545\,GHz CIB anisotropy level. The reason for this is that the 857\,GHz total flux map has been used to provide the dust optical depth at 353\,GHz and the CIB contamination at 857\,GHz is far less dominant than at 545 or 353\,GHz. This confirms that very little CIB contamination is present in the estimates of $\beta$ and $T$; the largest source of CIB contamination is from the optical depth. At 3000\,GHz any residual CIB anisotropies plus noise present in the optical depth parameter is extrapolated up to high frequencies following the MBB form and so increased in magnitude. Therefore the 3000\,GHz latitude plot shows absolute flux density differences slightly higher than the 3000\,GHz CIB anisotropies plus noise level across all latitudes. This could be avoided if another, cleaner, thermal dust template was chosen - for example the 3000\,GHz total flux map. The 857\,GHz total flux map was chosen in this work as only the 353, 545 and 857\,GHz simulated maps represent empirically taken data (the {\it{Planck}} HFI maps); the 3000\,GHz total flux simulation represents the IRIS combination of {\it{IRAS}} and {\it{COBE}}-DIRBE maps. 

\section{Conclusions}

Obtaining reliable thermal dust estimates from {\it{Planck}} HFI data is complicated by the presence of the CIB which ties the accuracy of any thermal dust estimates to the dust signal-to-noise ratio. We have presented a new method of parametric fitting based around informed, initial estimates of parameter values called {\texttt{premise}}. Full sky data are divided into patches of various size where each patch area contains only those pixels which share similar thermal dust properties. The MBB model is fit to each patch producing an all-sky estimate of the model parameters at a lower resolution than the raw data. These parameter estimates are then refined using a sparsity-based optimisation method to produce the final, full resolution parameter estimates. 

By comparing {\texttt{premise}} to a GNILC-like method over select regions of the sky we find that the success of the GNILC-like method is heavily dependent on the dust signal-to-noise ratio. {\texttt{premise}} can also be see to suffer from this effect, however, by taking advantage of the sparse nature of thermal dust, {\texttt{premise}} demonstrates an increasing ability to outperform the GNILC-like method at parameter estimation as signal-to-noise worsens. This is because instead of increasing the level of smoothing to compensate for poor signal-to-noise, {\texttt{premise}} increases the number of wavelet coefficients that are prevented from contributing to the reconstructed thermal dust estimate. Whereas smoothing unbiasedly removes all information (relating to both thermal dust and CIB) below a certain resolution, the wavelet thresholding instead only removes those coefficients associated with the CIB at all scales. The thresholding value is an important one as it is this value which decides which coefficients relate to CIB and which relate to thermal dust. This value is calculated at each wavelet scale. As thresholding in the wavelet domain allows us to detect any sparse emission above the noise level at all angular scales we do not suffer from the same loss of information at high resolution as those methods which make use of smoothing. Therefore we can present MBB parameter maps at full resolution. It should however be noted that if, for a particular patch of sky at small angular scales, the thermal dust signal is completely subdominant to the CIB then we will loose all information pertaining to thermal dust for that patch at that resolution. Additionally, by splitting the method into a fast initial estimation and then a refinement step, we introduce a considerable time reduction in the computational time required for {\texttt{premise}} to perform on full sky, $\rm{N_{side}}$ 2048 {\texttt{HEALPix}} maps. 

In this work we fit for the thermal dust temperature, optical depth at 353\,GHz and spectral index and produce all-sky maps of these three values with median absolute percentage deviations from the true simulation values of 1.7, 3.7 and 3.4 per cent, respectively. We use our dust parameters to reconstruct thermal dust estimates at 353, 545, 857 and 3000\,GHz with median absolute percentage deviations from the true simulation values of 5.1, 3.4 and 2.5 and 4.5 per cent, respectively.

The main limitation of {\texttt{premise}} is accurate parameter estimation within the very centre of the Galactic plane (due to a loss of thermal dust information after point source masking).  

\section*{Acknowledgements}
This work is supported by the European Community through the grant LENA (ERC StG no. 678282) within the H2020 Framework Program.



\bibliographystyle{mnras}
\bibliography{refs} 



\newpage

\appendix

\section{Large-scale CIB offset}\label{sec:apA}

To determine the large-scale CIB offset for each frequency, data from the Leiden/Argentine/Bonn (LAB) $\rm{H_{I}}$ survey \citep{lab} were used. The LAB data are integrated over radial velocity so three maps are available: the low velocity map (-30 to 30 km $\rm{s}^{-1}$), the intermediate velocity map (-100 to -30 km $\rm{s}^{-1}$) and the high velocity map (-500 to -100 km $\rm{s}^{-1}$). 

The total flux (353, 545, 857 and 3000\,GHz) and the LAB low and intermediate velocity all-sky maps were all downgraded to $\rm{N_{side}}$ 128 and smoothed to $1^{\circ}$. Two masks were then made which selected pixels for use depending on their column densities ($N_{\rm{H_{I}}}$) within the low and intermediate velocity maps :

\begin{tabular}{ l | c | c  }
\hline
  Mask & Low $N_{\rm{H_{I}}}$ ($\rm{cm}^{-2}$) & Intermediate $N_{\rm{H_{I}}}$ ($\rm{cm}^{-2}$)  \\
  \hline
  1 &  $< 2 \times 10^{20}$  & $ < 0.1 \times 10^{20}$ \\
  2 &  $ < 3 \times 10^{20}$  & -- \\
  \hline
\end{tabular}

Linear regression was used to determine the large-scale CIB offset at each frequency using the following masked data: 

\begin{tabular}{ l | c | c  }
\hline
 x axis data & y axis data & Mask applied \\
  \hline
  3000\,GHz & low + intermediate LAB & 2 \\ 
  857\,GHz &  low + intermediate LAB & 1 \\
  545\,GHz & 857\,GHz - $\rm{offset}_{857}$ & 1 \\
  353\,GHz & 857\,GHz - $\rm{offset}_{857}$  & 1 \\
  \hline
\end{tabular}

Fig.~\ref{fig:ciboffsetLinear} shows the linear regression between the 545\,GHz and the 857\,GHz total flux binned, mask 1 pixels. The 857\,GHz offset has been removed from the 857\,GHz data for the linear regression. 

\begin{figure}
\centering
\includegraphics[width=0.95\linewidth]{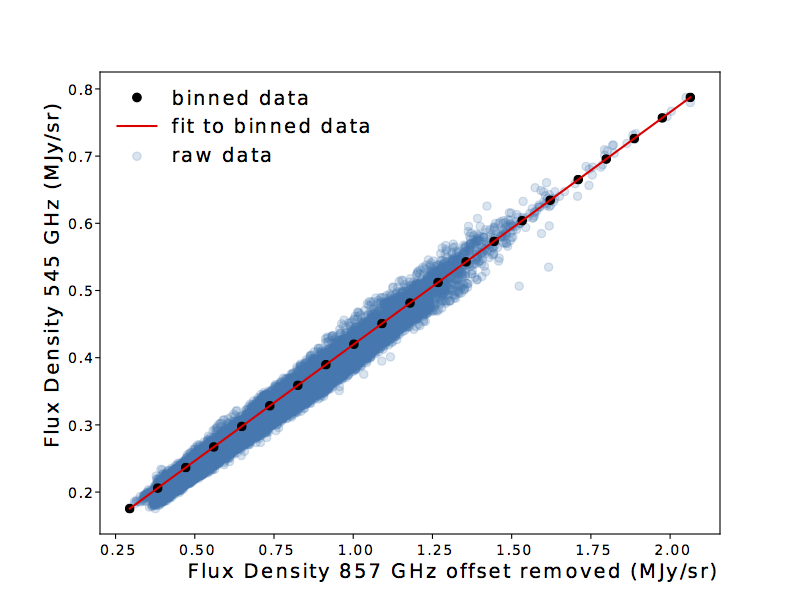}
\caption{An example of the linear regression used to determine the large scale CIB offset for each frequency.}
\label{fig:ciboffsetLinear}
\end{figure}

\section{GNILC-like methodology}\label{sec:apB}

The GNILC-like algorithm analysed in this work is based on that detailed in \citet{gnilc} and \citet{gnilc11}. Our implementation is described here in full so that any differences may be highlighted. Instead of working on the sphere and using the needlet transform, the total flux maps were divided into twelve 2048X2048 pixel faces and the wavelet transform was used on the 2D faces. The optimum number of wavelet scales was empirically found to be six, instead of the ten scales used by \citet{gnilc}. This difference can be explained by the fact that \citet{gnilc} work on all 12X2048X2048 pixels at once and the number of wavelet scales required is proportional to the log of the number of data samples. 

The following steps were completed for each wavelet scale ($j$):
\begin{enumerate}[label*=\arabic*.]
 \item The total flux maps were smoothed within the wavelet domain via convolution with a Gaussian of FWHM $2^{j} \times \frac{16\, \rm{pixels}}{2 \sqrt{2 \log{2}}}$. This width may be different from that used by \citet{gnilc}.  
 \item The nuisance terms were collected as N = CIB + CMB + instrumental noise. As simulation data were used the CIB, CMB and instrumental noise estimates were in fact the actual CIB, CMB and instrumental noise contributions to the total flux. The $\rm{N_{obs}}$ by $\rm{N_{obs}}$ nuisance covariance matrix was calculated as:
\begin{equation}
{\bf{R}}_{{\rm{nus}}} = \frac{1}{\rm{Npix}} \left({\bf{N \times N^{T} }} \right),  
\end{equation}     
 \item The covariance matrices of the smoothed total flux maps were calculated as $X_{\nu_{i}} \times X_{\nu_{f}}^{T} $ where $X_{\nu_{i/f}}$ is the total flux within the wavelet domain at frequency $i/f$.
\item The $\rm{N_{obs}}$ by $\rm{N_{obs}}$ total flux covariance matrices for each smoothed pixel were whitened: ${\bf{R}}_{{\rm{nus}}}^{-1/2} {\bf{R}}_{{\rm{tot}}} {\bf{R}}_{{\rm{nus}}}^{-1/2}$.
\item The eigenvectors of each whitened covariance matrix were ordered and the Akaike Information Criterion was used to select eigenvalues which deviated significantly from unity. Those eigenvectors (${\bf{U}}_{s}$) gave the mixing matrix (${\bf{F}} = {\bf{R}}_{{\rm{nus}}}^{1/2} {\bf{U}}_{s} $) used to obtain the leat-squares optimisation of thermal dust emission ${\bf{F}} \left( {\bf{F}}^{T} {\bf{R}}_{{\rm{tot}}}^{-1} {\bf{F}} \right) {\bf{F}}^{T} {\bf{R}}_{{\rm{tot}}}^{-1} {\bf{X}}$ 
\item If no eigenvectors were identified as significant then the total flux was believed to be dominated by the nuisance terms and so the signal contribution at that pixel and wavelet scale was masked out (set to zero). 
\end{enumerate}
The thermal dust contributions at each wavelet scale were recomposed to form the GNILC-like estimate of thermal dust emission for each frequency within pixel space. 


\bsp	
\label{lastpage}
\end{document}